\documentclass[pre,twocolumn,aps,10pt]{revtex4-2}
\usepackage{amsmath}
\usepackage{amssymb}
\usepackage{graphicx}
\usepackage{braket}
\usepackage{stmaryrd}
\usepackage{hyperref}
\usepackage{color}
\usepackage{dsfont}
\usepackage{bm}

\hypersetup{
    colorlinks,
    citecolor=blue,
    linkcolor=blue,
    urlcolor=blue
}

\begin{document}

\title{Thermodynamic Entropic Uncertainty Relation}
\author{Yoshihiko Hasegawa}
\email{hasegawa@biom.t.u-tokyo.ac.jp}
\affiliation{Department of Information and Communication Engineering, Graduate
School of Information Science and Technology, The University of Tokyo,
Tokyo 113-8656, Japan}

\author{Tomohiro Nishiyama}
\email{htam0ybboh@gmail.com}
\affiliation{Independent Researcher, Tokyo 206-0003, Japan}

\date{\today}
\begin{abstract}

Thermodynamic uncertainty relations reveal a fundamental trade-off between the precision of a trajectory observable and entropy production,
where the uncertainty of the observable is quantified by its variance.
In information theory, Shannon entropy is a common measure of uncertainty. However, a clear quantitative relationship between the Shannon entropy of an observable and the entropy production in stochastic thermodynamics remains to be established.
In this Letter, we show that an uncertainty relation can be formulated in terms of the Shannon entropy of an observable and the entropy production. We introduce symmetry entropy, an entropy measure that quantifies the symmetry of the observable distribution, and demonstrate that a greater asymmetry in the observable distribution requires higher entropy production.
Specifically, we establish that the sum of the entropy production and the symmetry entropy cannot be less than $\ln 2$.
As a corollary, we also prove that the sum of the entropy production and the Shannon entropy of the observable is no less than $\ln 2$.
As an application, we demonstrate our relation in the diffusion decision model, revealing a fundamental trade-off between decision accuracy and  entropy production in stochastic decision-making processes.

\end{abstract}
\maketitle

\section{Introduction\label{sec:Introduction}}

The thermodynamic uncertainty relation \cite{Barato:2015:UncRel,Gingrich:2016:TUP,Garrahan:2017:TUR,Dechant:2018:TUR,Terlizzi:2019:KUR,Hasegawa:2019:CRI,Hasegawa:2019:FTUR,Dechant:2020:FRIPNAS,Vo:2020:TURCSLPRE,Koyuk:2020:TUR} expresses a universal trade-off between the entropy production and the relative variance of time-integrated observables for systems operating far from equilibrium. For a steady-state stochastic thermodynamic system and a trajectory observable $F$, the thermodynamic uncertainty relation states \cite{Barato:2015:UncRel,Gingrich:2016:TUP}
\begin{align}
    \frac{\mathrm{Var}[F]}{\mathbb{E}[F]^{2}} \geq \frac{2}{\Sigma},
    \label{eq:TUR_def}
\end{align}
where $\mathbb{E}[F]$ and $\mathrm{Var}[F]$ are the expectation value and variance of $F$, respectively, and $\Sigma$ denotes the entropy production during the observation interval. 
The thermodynamic uncertainty relation implies that achieving higher precision inevitably requires greater entropy production, a manifestation of the ``no free lunch'' principle in thermodynamics.
The thermodynamic uncertainty relation given by Eq.~\eqref{eq:TUR_def} quantifies the uncertainty of the observables via their variance;
however, uncertainty is often evaluated using the Shannon entropy in the context of information theory. 
Consider a thermodynamic decision-making process that yields a categorical output, such as a ``yes'' or ``no'' label. Since the output is categorical rather than numerical, we cannot use variance to quantify its uncertainty.
However, even in such cases, we can still calculate the Shannon entropy. Therefore, to analyze the trade-off between the accuracy of the categorical outputs and the entropy production, a thermodynamic uncertainty relation based on Shannon entropy is required.

The issue of how to quantify uncertainty has been actively studied in quantum mechanics.
In the Heisenberg-Robertson uncertainty relation \cite{Heisenberg:1927:UR,Robertson:1929:UncRel}, the uncertainty in the outcomes of two observables is quantified through their variances. 
Later, Hirschman showed that the variance can be replaced by the Shannon entropy \cite{Hirschman:1957:EntropyUncertainty,BialynickiBirula:1975:EUR}. 
Maassen and Uffink \cite{Maassen:1988:Entropic} derived a precise statement of this entropic uncertainty relation \cite{Coles:2017:EntropicUR}; the sum of the entropies of the two observables cannot fall below a bound determined by the extent of their eigenstate overlap. 
The entropic uncertainty relation plays a fundamental role in quantum cryptography, especially in quantum key distribution protocols \cite{Berta:2010:MemoryEUR}.

In this Letter, in the context of stochastic thermodynamics, we show that there is an uncertainty relation between entropy production and the \textit{symmetry entropy} [cf. Eq.~\eqref{eq:R_def}], which quantifies the extent of symmetry of the observable's distribution.
Specifically, we show that the sum of the entropy production and the symmetry entropy of the observable should be no less than $\ln 2$ [cf. Eqs.~\eqref{eq:main_result1} and \eqref{eq:main_result2}]. 
In other words, the asymmetry of the observable probability distribution, as quantified by entropy, requires that the entropy production be at least equal to this measure of asymmetry.
As a corollary of the result, we also show that the sum of the entropy production and the Shannon entropy of the observable should be no less than $\ln 2$
[Eq.~\eqref{eq:main_result_binary}].
We apply the derived relation to the diffusion decision model, illustrating a trade-off between decision accuracy and entropy production in stochastic decision-making processes.
The trade-off relation in thermodynamic cost has been actively studied, particularly in terms of thermodynamic uncertainty relations \cite{Barato:2015:UncRel,Gingrich:2016:TUP,Garrahan:2017:TUR,Dechant:2018:TUR,Terlizzi:2019:KUR,Hasegawa:2019:CRI,Hasegawa:2019:FTUR,Dechant:2020:FRIPNAS,Vo:2020:TURCSLPRE,Koyuk:2020:TUR} using the variance of observable quantities and speed limits using the distance between states \cite{Shiraishi:2018:SpeedLimit,Ito:2020:TimeTURPRX,Ito:2018:InfoGeo,Vo:2020:TURCSLPRE,Vu:2021:GeomBound,Dechant:2019:Wasserstein,Vu:2022:OptimalTransportPRX}. This study demonstrates the trade-off between the entropy of the observable and the entropy production, which is expected to lead to the derivation of other trade-off relations.

\begin{figure}
    \includegraphics[width=0.9\linewidth]{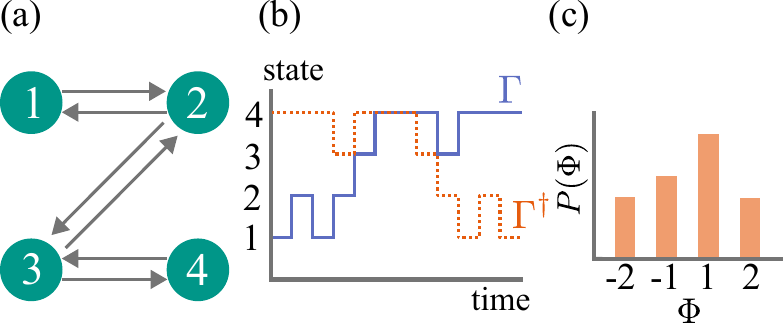}
    \caption{
    Conceptual representation of the thermodynamic entropic uncertainty relation.
    (a) Stochastic thermodynamic process. The thermodynamic entropic uncertainty relation considers a stochastic process, where each state transition is random. 
    (b) 
    Trajectory of the stochastic process shown in (a).
    $\Gamma$ denotes the time evolution of a realization of the process. 
    $\Gamma^\dagger$ is the time-reversal of $\Gamma$. 
    (c)
    Probability distribution of observable $\Phi(\Gamma)$.
    $\Phi(\Gamma)$ is arbitrary, provided it satisfies the time-reversal antisymmetric property [Eq.~\eqref{eq:time_reversal}]. 
    }
    \label{fig:ponch}
\end{figure}

\begin{figure}
    \includegraphics[width=0.99\linewidth]{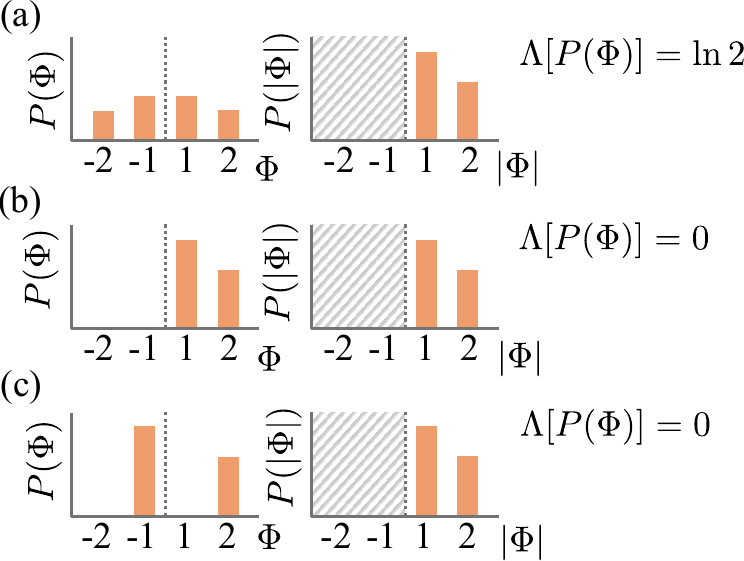}
    \caption{
    Examples of the symmetry entropy $\Lambda[P(\Phi)]$. Horizontal axes denote values of $\Phi$ (left column) and $|\Phi|$ (right column). The values of $\Phi$ always form pairs; $-1$ and $1$, and $-2$ and $2$ are two pairs in the examples. 
    Vertical axes denote the probability distribution $P(\Phi)$ (left column) and $P(|\Phi|)$ (right column). 
    (a)  Distribution symmetric with respect to $\Phi=0$, where the probabilities of all of the pairs are identical. 
    $P(|\Phi|)$ is different from the original distribution $P(\Phi)$ and thus $\Lambda[P(\Phi)]$ is $\ln 2$. 
    (b) Asymmetric distribution where the probabilities of all of the pairs are fully biased.
    $P(|\Phi|)$ and $P(\Phi)$ are identical and thus $\Lambda[P(\Phi)]$ is $0$. 
    (c) Asymmetric distribution where probabilities of all of the pairs are fully biased.
    The distributions $P(|\Phi|)$ and $P(\Phi)$ are different, but they effectively become the same if the labels for the pair $-1$ and $1$ are swapped. Therefore, they are essentially the same distribution. This results in $\Lambda[P(\Phi)]=0$.
}
    \label{fig:R_expl}
\end{figure}

\section{Methods\label{sec:Methods}}

Let $X$ be a random variable and $P(X)$ be its probability distribution. 
The Shannon entropy $H[P(X)]$ is defined by
\begin{align}
    H[P(X)]\equiv-\sum_{x}P(X=x)\ln P(X=x).
    \label{eq:Shannon_def}
\end{align}
Let $Y$ be another random variable.
The Kullback-Leibler divergence between $P(X)$ and $P(Y)$ is defined by
\begin{align}
    D[P(X)\|P(Y)]&\equiv\sum_{x}P(X=x)\ln\frac{P(X=x)}{P(Y=x)}\nonumber\\&=-H[P(X)]+C[P(X),P(Y)],
    \label{eq:KL_def}
\end{align}
where $C[P(X),P(Y)]$ is the cross entropy:
\begin{align}
    C[P(X),P(Y)]\equiv-\sum_{x}P(X=x)\ln P(Y=x).
    \label{eq:cross_entropy_def}
\end{align}
It is well known that the Kullback-Leibler divergence is non-negative. 
Moreover, the Kullback-Leibler divergence satisfies monotonicity. 
Consider a transformation that maps the original random variables $X$ and $Y$ to new random variables $\tilde{X}$ and $\tilde{Y}$, respectively.
Then the following monotonicity relation holds:
\begin{align}
    D[P(X)\|P(Y)]\ge D[P(\tilde{X})\|P(\tilde{Y})],
    \label{eq:KL_monotonicity}
\end{align}
where $P(\tilde{X})$ and $P(\tilde{Y})$ are probability distributions of the transformed variables $\tilde{X}$ and $\tilde{Y}$, respectively.

Having introduced basic concepts of the divergence, we move to consider stochastic thermodynamic systems. 
Stochastic thermodynamics considers processes whose state changes are described by a stochastic process (Fig.~\ref{fig:ponch}(a)). 
Let $\Gamma$ be a stochastic trajectory of the process and $\Gamma^\dagger$ be its time reversal (Fig.~\ref{fig:ponch}(b)). 
Moreover, we can define the probability of measuring $\Gamma$, which is denoted by $\mathcal{P}(\Gamma)$. 
Assuming the local detailed balance, it is known that  the entropy production under the steady-state condition is defined by the Kullback-Leibler divergence \cite{Crooks:1999:CFT,Seifert:2005:FT}:
\begin{align}
    \Sigma=D[\mathcal{P}(\Gamma)\|\mathcal{P}(\Gamma^{\dagger})].
    \label{eq:entropy_production}
\end{align}
Note that the expression in Eq.~\eqref{eq:entropy_production} also holds for continuous processes such as Langevin dynamics.
Consider an observable $\Phi(\Gamma)$, which is a function of the trajectory $\Gamma$.
Here, we assume that
$\Phi(\Gamma)$ is antisymmetric under the time reversal: 
\begin{align}
    \Phi(\Gamma) = -\Phi(\Gamma^\dagger).
    \label{eq:time_reversal}
\end{align}
For example, $\Phi(\Gamma)$ represents a thermodynamic current. Important thermodynamic quantities, such as stochastic dissipated heat or displacement, are expressed by $\Phi(\Gamma)$.
In the supplementary material \cite{Supp:2025:TEUR}, we provide a generalized formulation where the time-reversal antisymmetric condition [Eq.~\eqref{eq:time_reversal}] is relaxed. 
Initially, we assume that $\Phi(\Gamma)$ takes values in a countable set, that is, the probability distribution $P(\Phi)$ is discrete (Fig.~\ref{fig:ponch}(c)).  
However, most of the results below hold for the continuous case as well. 
\nocite{Boyd:2004:ConvexOptimBook}

\section{Results\label{sec:Results}}

We first introduce an entropic measure that quantifies a trajectory observable.
Let $\Lambda[P(\Phi)]$ be
\begin{align}
    \Lambda[P(\Phi)] \equiv H[P(\Phi)]-H[P(|\Phi|)],
    \label{eq:R_def}
\end{align}
which
is the entropy difference between the original distribution $P(\Phi)$ and its absolute-value distribution $P(|\Phi|)$. 
Here, we call $\Lambda[P(\Phi)]$ \textit{symmetry entropy}. 
The symmetry entropy $\Lambda[P(\Phi)]$ quantifies the extent of symmetry of $P(\Phi)$. 
Intuitively speaking, the difference $H[P(\Phi)]-H[P(|\Phi|)]$ quantifies how much ``sign'' information remains in the distribution $P(\Phi)$ once its magnitude $P(|\Phi|)$ is known.

For simplicity, let us consider the case where the observable $\Phi(\Gamma)$ does not take the value $0$, a condition which holds in several problem settings.
For instance, we may consider binary classification based on trajectories of stochastic processes.
In this case, the observable $\Phi(\Gamma)$ does not include $0$. 
$\Phi(\Gamma)=0$ should be handled separately because $\Phi(\Gamma) = \Phi(\Gamma^\dagger) = 0$ from Eq.~\eqref{eq:time_reversal}, showing that the observable is invariant under the time reversal. 
Later, we will consider the case where $\Phi(\Gamma) = 0$ is included.
It can be shown that 
\begin{align}
    0\le \Lambda[P(\Phi)] \le \ln 2,
    \label{eq:R_range}
\end{align}
whose proof is provided in 
Appendix~\ref{sec:derivation_R_range}.
Let us elaborate on the necessary and sufficient conditions for achieving $0$ and $\ln 2$ in Eq.~\eqref{eq:R_range}. 
Figure~\ref{fig:R_expl} illustrates examples of the symmetry entropy, with the horizontal axes representing the values of $\Phi$ (left column) and $|\Phi|$ (right column), and the vertical axes corresponding to $P(\Phi)$ (left column) and $P(|\Phi|)$ (right column).
From the time-reversal antisymmetric condition [Eq.~\eqref{eq:time_reversal}], if $\Phi = \phi$ exists ($\phi>0$), then $\Phi = -\phi$ also exists, which is regarded as a pair of the observable. 
From the calculation in Appendix~\ref{sec:derivation_R_range}, $\Lambda[P(\Phi)] = \ln 2$ if and only if $P(\Phi=\phi) = P(\Phi = -\phi)$ for all $\phi$. 
This corresponds to the symmetric probability distribution as shown in Fig.~\ref{fig:R_expl}(a). 
In contrast, $\Lambda[P(\Phi)] = 0$ is satisfied if and only if $P(\Phi=\phi) = 0$ or $P(\Phi=-\phi) = 0$ for all $\phi$. 
This case corresponds to asymmetric probability distributions as shown in Figs.~\ref{fig:R_expl}(b) and (c). 
When $\Phi(\Gamma)$ is a binary function (outputs two categories), we have $\Lambda[P(\Phi)] = H[P(\Phi)]$.

Using the symmetry entropy $\Lambda[P(\Phi)]$, we obtain the trade-off between the entropy production and the asymmetry of $P(\Phi)$ quantified by entropy:
\begin{align}
    \Sigma\ge\ln2-\Lambda[P(\Phi)]\ge0.
    \label{eq:main_result1}
\end{align}
Equation~\eqref{eq:main_result1} is the main result of this study and is referred to as the \textit{thermodynamic entropic uncertainty relation}. 
The derivation is shown in 
Appendix~\ref{sec:derivation_main}.
The right-hand side of Eq.~\eqref{eq:main_result1} quantifies the asymmetry of the probability distribution $P(\Phi)$. 
Equation~\eqref{eq:main_result1} demonstrates that for any observable $\Phi$ that satisfies the time-reversal antisymmetric condition specified in Eq.~\eqref{eq:time_reversal}, the entropy production must be at least $\ln 2 - \Lambda[P(\Phi)]$.
The trade-off between entropy production and observable asymmetry parallels traditional thermodynamic uncertainty relations, which illustrate a trade-off between the relative variance of observables and entropy production. 
The thermodynamic uncertainty relations consider
the relative variance $\mathrm{Var}[\Phi]/\mathbb{E}[\Phi]^{2}$. 
In a sense, using the relative variance can also be seen as quantifying the asymmetry of $P(\Phi)$;
when the variance is smaller and the expectation is greater, the probability distribution $P(\Phi)$ is more asymmetric with respect to $\Phi=0$. 
Since $\ln 2 - \Lambda[P(\Phi)] \ge 0$ in Eq.~\eqref{eq:main_result1} due to Eq.~\eqref{eq:R_range}, Eq.~\eqref{eq:main_result1} can be regarded as a refinement of the second law, employing the entropy of the observable. 
There are some advantages in employing the entropy instead of the variance. 
When the observable is a categorical variable that outputs labels such as ``yes'' or ``no'', which is the case in classification tasks, the use of variance is not appropriate.
Even in such cases, 
the symmetry entropy defined by Eq.~\eqref{eq:R_def} can handle categorical outputs.
As mentioned, the outputs of the observables constitute pairs, i.e., $\Phi=\phi$ and $\Phi=-\phi$.
As the value of $\phi$ does not affect the calculation of the entropy $H[P(\Phi)]$, the paired outputs can be replaced with categorical outputs such as $\Phi=\phi = \text{``yes''}$ or $\Phi=-\phi= \text{``no''}$. 
The right-hand side of Eq.~\eqref{eq:main_result1} can also be interpreted as a divergence. If we denote the Jensen-Shannon divergence by $\mathrm{JS}[P(X)\|P(Y)]$, then
\begin{align}
    \mathrm{JS}[P(\Phi)\|P(-\Phi)] = \ln 2 - \Lambda[P(\Phi)],
    \label{eq:rhs_JS}
\end{align}
the derivation of which is shown in 
Appendix~\ref{sec:symmetry_entropy_Jensen_Shannon}.

Equation~\eqref{eq:main_result1} provides a trade-off between the symmetry entropy and the entropy production.
For a discrete probability distribution, $H[P(|\Phi|)] \ge 0$ holds.
Therefore, $H[P(\Phi)] \ge \Lambda[P(\Phi)]$ and thus
the following bound also holds:
\begin{align}
    \Sigma \ge \ln 2 - H[P(\Phi)],
    \label{eq:main_result_binary}
\end{align}
which purely relates the entropy production $\Sigma$ and the Shannon entropy of the observable $\Phi$. 
The right-hand side of Eq.~\eqref{eq:main_result_binary} may not always be non-negative.
Although the form of Eq.~\eqref{eq:main_result_binary} is more appealing in terms of physical interpretation, the bound is weaker.
When the observable $\Phi(\Gamma)$ is a binary function, the right-hand side of Eq.~\eqref{eq:main_result_binary} becomes non-negative as $H[P(\Phi)] = \Lambda[P(\Phi)]$ for the binary case.

So far, we assumed that $\Phi(\Gamma)$ takes values in a countable set, that is, $P(\Phi)$ is a discrete distribution. 
Consider the case where $\Phi(\Gamma)$ produces continuous values, where the summation should be replaced by an integration. 
When considering a continuous distribution, the notable difference is that the differential entropy may take negative values. 
However, the existence of negative values is not problematic, as such negative values are offset in $H[P(\Phi)] - H[P(|\Phi|)]$. 
Therefore, the main result of Eq.~\eqref{eq:main_result1} holds for the continuous case as well. 

So far, we assumed that the observable $\Phi(\Gamma)$ does not include $\Phi(\Gamma)=0$.
It is straightforward to extend the result to the case where the observable includes $\Phi(\Gamma)=0$. 
Specifically, when $\Phi(\Gamma)$ can equal $0$, the range of $\Lambda[P(\Phi)]$ is modified as follows (see Appendix~\ref{sec:derivation_R_range}):
\begin{align}
    0\le \Lambda[P(\Phi)]\le[1-P(\Phi=0)]\ln2.
    \label{eq:R_range2}
\end{align}
The bound becomes
\begin{align}
    \Sigma\ge[1-P(\Phi=0)]\ln2-\Lambda[P(\Phi)]\ge 0,
\label{eq:main_result2}
\end{align}
which includes Eq.~\eqref{eq:main_result1} as the specific case $P(\Phi=0)=0$. 
The derivation is shown in Appendix~\ref{sec:derivation_main}.
As long as the entropy production is given by Eq.~\eqref{eq:entropy_production},
Eq.~\eqref{eq:main_result2} holds for any observable $\Phi(\Gamma)$ satisfying Eq.~\eqref{eq:time_reversal}.
In Eq.~\eqref{eq:main_result2}, $\Phi = 0$ plays a special role. 
When considering the process of doing nothing, the observable $\Phi$ is always $0$, implying $P(\Phi=0)=1$. 
For such empty dynamics, the entropy production is $0$, showing that both sides of Eq.~\eqref{eq:main_result2} equal $0$ and thus the inequality becomes an equality.
Although the conventional thermodynamic uncertainty relation [Eq.~\eqref{eq:TUR_def}] does not handle $\Phi=0$ as an exceptional case,
it becomes ill-defined for the equilibrium condition where both the expectation of the current and the entropy production vanish. 
Equations~\eqref{eq:main_result1} and \eqref{eq:main_result2} saturate if and only if the system is in equilibrium.
For a detailed discussion on the saturation condition, see 
Appendix~\ref{sec:equality_condition}.
Again, when $\Phi(\Gamma)$ is a discrete distribution, the following relation holds as well:
\begin{align}
    \Sigma \geq[1-P(\Phi=0)] \ln 2-H[P(\Phi)].
    \label{eq:TEUR_binary_P0}
\end{align}

We remark on the relation between Eq.~\eqref{eq:main_result1} and the Landauer principle \cite{Landauer:1961:LP}.
Consider a process that resets a system, which has an equal probability of being in one of two states, to a single specific state. The following relation is obtained:
\begin{align}
    \Delta S_\mathrm{m} \ge \ln 2,
    \label{eq:Landauer}
\end{align}
where $\Delta S_\mathrm{m}$ is the entropy increase in the surrounding medium. 
Note that the Landauer principle can be considered as a specific case of the second law of thermodynamics \cite{Reed:2014:Landauer}. 
Given the assumption that the system is in steady state, the change in system entropy $\Delta S$ is zero; consequently, $\Sigma = \Delta S_\mathrm{m}$.
Then, Eq.~\eqref{eq:main_result1} can be expressed by $\Delta S_{\mathrm{m}}\ge\ln2-\Lambda[P(\Phi)]$.
When the probability distribution $P(\Phi)$ is totally asymmetric, i.e., $\Lambda[P(\Phi)]=0$, Eq.~\eqref{eq:main_result1} is formally identical to the Landauer principle. 
However, note that the right-hand side of Eq.~\eqref{eq:Landauer} arises from a reduction in the Shannon entropy within the system's state, implying that the Landauer principle is relevant when the system state changes over the time evolution.
This contrasts with Eq.~\eqref{eq:main_result1}, where the system is assumed to be in steady state, and $\ln 2$ on its right-hand side arises from the asymmetry of the observable, not the state of the system. 

\section{Example\label{sec:Example}}

\begin{figure}
\centering
\includegraphics[width=0.8\linewidth]{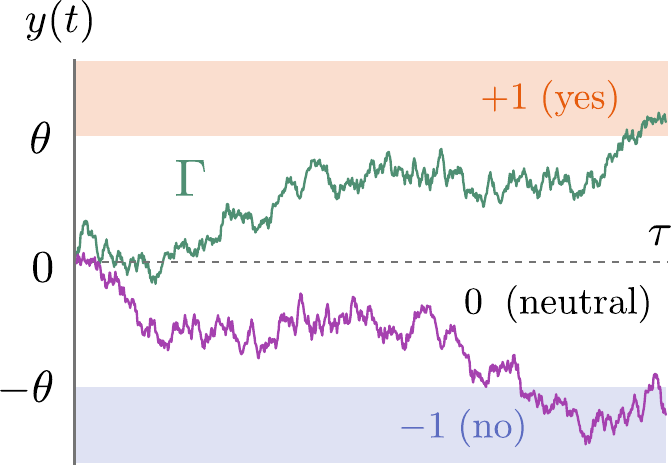}
\caption{Description of the diffusion decision model. The green trajectory illustrates a realization that exceeds the threshold $\theta$, resulting in a decision of $+1$ (``yes'' class). Conversely, the purple trajectory represents a realization that falls below $-\theta$, leading to a decision of $-1$ (``no'' class).
Trajectories that do not fall into either category correspond to a decision of $0$ (``neutral'' class).
}
\label{fig:DDM_illustration}
\end{figure}

\begin{figure}
\centering
\includegraphics[width=0.8\linewidth]{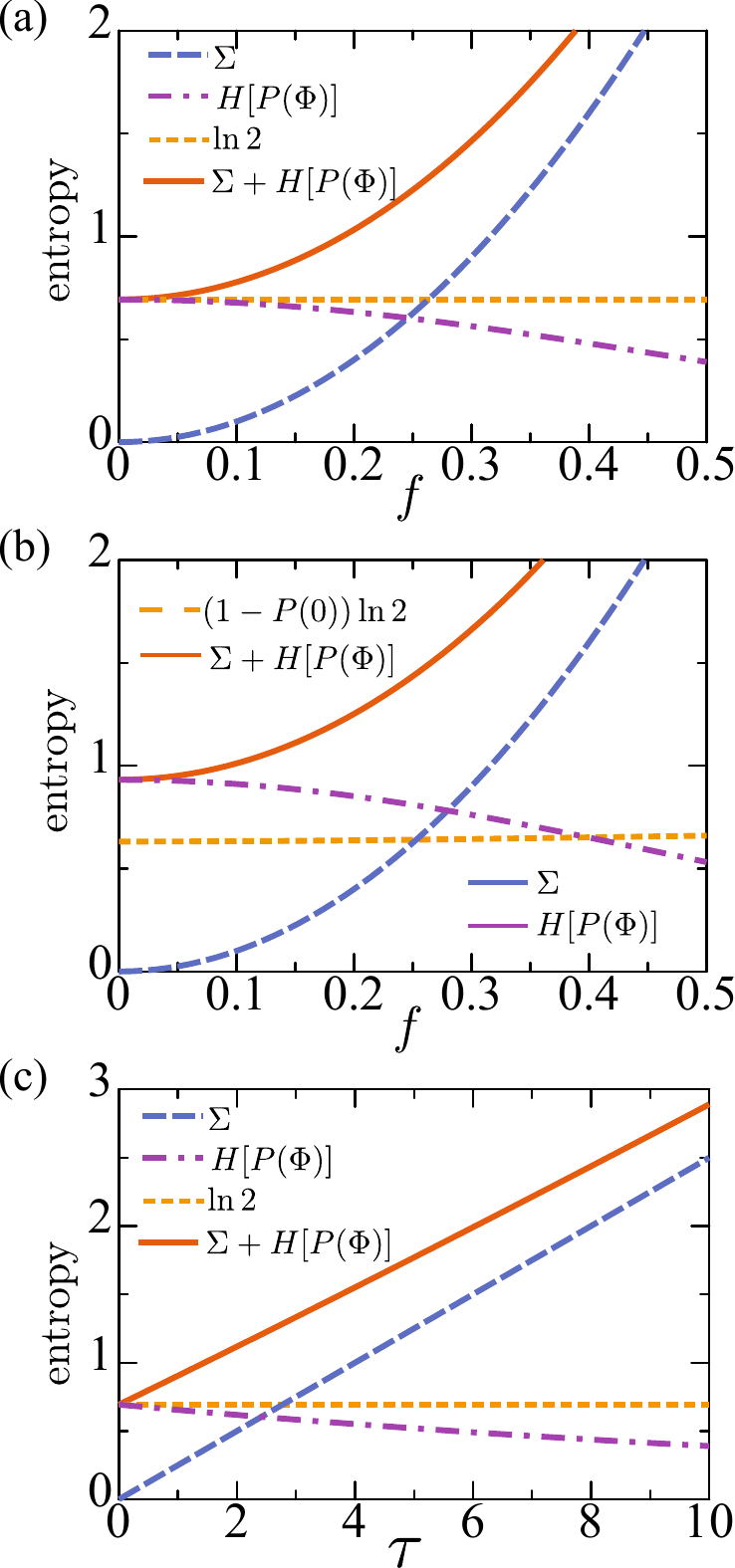}
\caption{
Results for the diffusion decision model with the discrete-output observable [Eq.~\eqref{eq:Phi_def_main}]. 
Panels (a) and (b) show entropic quantities as a function of parameter $f$
for $\theta = 0$ $\theta = 0.5$, respectively, with $A=1$ and $\tau=10$.
Panel (c) shows entropic quantities as a function of $\tau$ with $\theta = 0$, $A = 1$, and $f=0.5$. 
In these figures, the blue dashed line shows the entropy production $\Sigma$, and the purple dot-dashed line represents $H[P(\Phi)]$. Their sum, $\Sigma + H[P(\Phi)]$, is also depicted by the red solid line. The orange dotted line represents $\ln 2$ in (a) and (c) and $(1-P(0))\ln 2$ in (b), which serves as the lower bound for $\Sigma + H[P(\Phi)]$ according to Eq.~\eqref{eq:TEUR_binary_P0}. 
}
\label{fig:DDM_discrete_simulation}
\end{figure}

We demonstrate an application of the thermodynamic entropic uncertainty relation. Decision-making, a fundamental aspect of information processing, is crucial in diverse fields ranging from biology to information science. To illustrate this, we consider the diffusion decision model \cite{Ratcliff:2008:DDM,Ratcliff:2016:DDM,Myers:2022:DDM}, which describes decision-making as a stochastic process of evidence accumulation.
Let $\Gamma$ be a stochastic trajectory of the decision model within the time interval $[0,\tau]$ and $\Phi(\Gamma)$ be a function of $\Gamma$. 
In the diffusion decision model, a decision occurs when the accumulated evidence during $[0,\tau]$ surpasses thresholds.
We assume that if the evidence is greater than or equal to $\theta$, then $\Phi=+1$ (``yes'' class); if it is less than or equal to $-\theta$, then $\Phi=-1$ (``no'' class); and in all other cases, $\Phi=0$ (neutral class). 
Figure~\ref{fig:DDM_illustration} describes an example of the decision process. 
Here, $+1, -1, 0$ are simply labels, so we can assign other labels.
Specifically, let us consider the following Langevin dynamics for modeling the evidence accumulation often employed in the literature \cite{Ratcliff:2008:DDM,Ratcliff:2016:DDM,Myers:2022:DDM}:
\begin{align}
    \frac{dx}{dt}=f+\sqrt{2A}\xi(t),
    \label{eq:dzdt_def}
\end{align}
where $f \in \mathbb{R}$ is a constant force, $\xi(t)$ is the white Gaussian noise with the correlation $\braket{\xi(t)\xi(t^\prime)} = \delta(t-t^\prime)$, and $A$ is the noise intensity. 
We define the accumulated evidence as $y(\tau) = \int_0^\tau \partial_t x(t) dt$ and the observable is 
\begin{align}
\Phi(\Gamma)=\begin{cases}
+1\;\;(\text{yes}) & {\displaystyle y(\tau)\ge\theta}\\
-1\;\;(\text{no}) & {\displaystyle y(\tau)\le-\theta}\\
0\;\;(\text{neutral}) & \mathrm{otherwise}
\end{cases}.
\label{eq:Phi_def_main}
\end{align}
Explicit expressions of the probability distribution of $y$ are shown in Appendix~\ref{sec:DDM}. The entropy production during $[0,\tau]$ is given by
\begin{align}
\Sigma=\frac{\tau f^{2}}{A}.
    \label{eq:EP_def}
\end{align}
The thermodynamic entropic uncertainty relation loses its predictive power when the entropy production $\Sigma$ exceeds the threshold $\ln 2$. For the diffusion decision model, this occurs when the time parameter $\tau$ exceeds the critical value $\tau^*$ defined as
\begin{align}
    \tau^{*}\equiv\frac{A}{f^{2}}\ln2.
    \label{eq:tau_start_def}
\end{align}

Here, our interest is in the relationship between the entropy of the selected decision $H[P(\Phi)]$ and the entropy production $\Sigma$ of the Langevin equation.
Since $\Phi(\Gamma)$ satisfies the time-reversal antisymmetric condition [Eq.~\eqref{eq:time_reversal}], 
it satisfies Eqs.~\eqref{eq:main_result2} and \eqref{eq:TEUR_binary_P0},
implying that greater entropy production is necessary for lower decision entropy $H[P(\Phi)]$ under the steady-state condition.
In decision making, it is well known that there exists a speed-accuracy trade-off \cite{Stone:1960:ChoiceReaction,Zhang:2014:SpeedAccuracy}, responding faster often reduces accuracy, while achieving higher accuracy lengthens decision time.
Through the thermodynamic entropic uncertainty relation, we show quantitatively that another trade-off, an energy-accuracy trade-off, holds in the decision model.

We perform numerical calculations for the diffusion decision model [Eq.~\eqref{eq:dzdt_def}] to confirm Eq.~\eqref{eq:TEUR_binary_P0} \cite{Supp:2025:TEUR}. 
We calculate quantities, such as $\Sigma$ and $H[P(\Phi)]$, as a function of $f$ in Figs.~\ref{fig:DDM_discrete_simulation}(a) and (b) for different threshold settings (a) $\theta=0$ and (b) $\theta = 0.5$. 
We consider the steady-state condition. 
In Figs.~\ref{fig:DDM_discrete_simulation}(a) and (b), $\Sigma + H[P(\Phi)]$ is represented by the solid line, where the dotted line indicates (a) $\ln 2$ or (b) $(1-P(0))\ln 2$. 
As shown in Figs.~\ref{fig:DDM_discrete_simulation}(a) and (b), $\Sigma + H[P(\Phi)]$ is always above $\ln 2$ or $(1-P(0))\ln 2$, respectively, confirming the validity of Eq.~\eqref{eq:TEUR_binary_P0}.
It is clear that in the case of $\theta=0$, the values of $\Sigma + H[P(\Phi)]$ and the lower bound are closer than in the case of $\theta=0.5$. 
To examine how each of these components, $\Sigma$ and $H[P(\Phi)]$, behave as a function of $f$, $\Sigma$ and $H[P(\Phi)]$ are represented by the dashed line and the dot-dashed line, respectively.
Regarding entropy production $\Sigma$, it falls below $\ln 2$ in the region where $f$ ranges from $0.0$ to $0.25$ in Fig.~\ref{fig:DDM_discrete_simulation}(a). 
On the other hand, the entropy of the observable $H[P(\Phi)]$ is decreasing as a function of $f$, and except for $f = 0.0$, it is below $\ln 2$.
From these results, we can conclude that the inequality $\Sigma + H[P(\Phi)] \ge \ln 2$ holds only when considering the sum, $\Sigma + H[P(\Phi)]$, that is, $\Sigma$ or $H[P(\Phi)]$ alone falls below $\ln 2$. 

We study the $\tau$ dependence of the entropic quantities. 
We calculate the entropic quantities, such as $\Sigma$ and $H[P(\Phi)]$, as a function of $\tau$ in Fig.~\ref{fig:DDM_discrete_simulation}(c).
The meaning of each line is the same as in Fig.~\ref{fig:DDM_discrete_simulation}(a). 
Since the entropy production increases linearly as a function of $\tau$ [Eq.~\eqref{eq:EP_def}], 
we see that the
predictive power of the thermodynamic entropic uncertainty relation is lost around $\tau^* = 2.8$ [Eq.~\eqref{eq:tau_start_def}] for this parameter setting.

\begin{figure}
\centering
\includegraphics[width=0.8\linewidth]{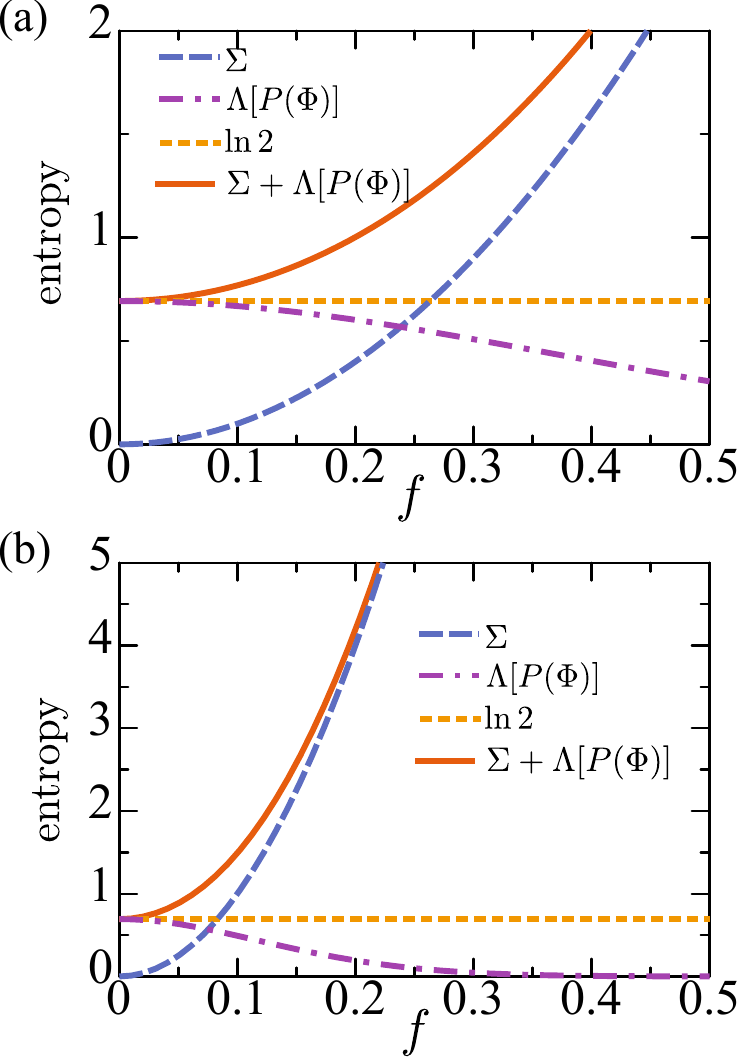}
\caption{
Results for the diffusion decision model with the continuous-output observable [Eq. \eqref{eq:Phi_continuous}]. Panels (a) and (b) show the entropic quantities versus $f$ for $A = 1$ and $A = 0.1$, respectively, with $\tau$ fixed at $10$.
The blue dashed line shows the entropy production $\Sigma$, and the purple dot-dashed line represents $\Lambda[P(\Phi)]$, the symmetry entropy. Their sum, $\Sigma + \Lambda[P(\Phi)]$, is also depicted with the red solid line. The orange dotted line displays $\ln 2$, which serves as the lower bound for $\Sigma + \Lambda[P(\Phi)]$ according to Eq.~\eqref{eq:main_result2}. 
}
\label{fig:DDM_continuous_simulation}
\end{figure}

Next, we consider the continuous-output observable case,
where the observable is 
\begin{align}
    \Phi(\Gamma)=y(\tau).
    \label{eq:Phi_continuous}
\end{align}
Equation~\eqref{eq:Phi_continuous} obviously satisfies the time-reversal antisymmetry condition. 
We numerically calculate the entropic quantities.
We show $\Sigma$ and $\Lambda[P(\Phi)]$ as a function of $f$ in Fig.~\ref{fig:DDM_continuous_simulation} for different noise intensity settings (a) $A=1$ and (b) $A = 0.1$. 
Figure~\ref{fig:DDM_continuous_simulation} shows the entropic quantities: $\Sigma + \Lambda[P(\Phi)]$ (solid line), $\ln 2$ (dotted line), $\Lambda[P(\Phi)]$ (dot-dashed line), and $\Sigma$ (dashed line).
Again, Figs.~\ref{fig:DDM_continuous_simulation}(a) and (b) demonstrate that the sum $\Sigma + \Lambda[P(\Phi)]$ consistently exceeds $\ln 2$, supporting the correctness of Eq.~\eqref{eq:main_result2}. 
Specifically, when the noise intensity $A$ is set to the smaller value, as shown in Fig.~\ref{fig:DDM_continuous_simulation}(b), the symmetry entropy $\Lambda[P(\Phi)]$ approaches zero as $f$ increases. 
When $A$ is smaller, the entropy production $\Sigma$ increases, leading to a rapid increase in the value of $\Sigma + \Lambda[P(\Phi)]$ as a function of $f$.
In Fig.~\ref{fig:DDM_continuous_simulation}(a),
in the range where $f$ varies from $0.0$ to $0.25$, the entropy production $\Sigma$ is less than $\ln 2$. The symmetry entropy $\Lambda[P(\Phi)]$ decreases with increasing $f$ and remains below $\ln 2$ for all values except when $f = 0.0$. 
Based on these observations, it is evident that the inequality $\Sigma + \Lambda[P(\Phi)] \ge \ln 2$ is satisfied only when considering the combined terms, $\Sigma + \Lambda[P(\Phi)]$. Individually, both $\Sigma$ and $\Lambda[P(\Phi)]$ drop below $\ln 2$.
\section{Conclusion\label{sec:Conclusion}}

In this study, we established the thermodynamic entropic uncertainty relation that links the entropy production and the Shannon entropy of the observable. Our findings extend conventional thermodynamic uncertainty relations by incorporating measures based on entropy, highlighting the role of Shannon entropy of observables in stochastic thermodynamics. The derived inequality formalizes a fundamental trade-off between entropy production and the asymmetry of the observable distribution. 
Through the application to the diffusion decision model, we revealed that accurate decisions necessarily require higher entropy production.
This framework provides a deeper understanding of nonequilibrium thermodynamics and expands the application of entropy-based uncertainty relations in stochastic systems. 
We have considered the Kullback-Leibler divergence with respect to trajectories $\Gamma$, but it is also possible to start from the divergence between different quantities. 
Another direction of expansion is towards quantum systems. In recent years, the thermodynamic uncertainty relations in quantum systems \cite{Erker:2017:QClockTUR,Brandner:2018:Transport,Carollo:2019:QuantumLDP,Liu:2019:QTUR,Guarnieri:2019:QTURPRR,Saryal:2019:TUR,Hasegawa:2020:QTURPRL,Hasegawa:2020:TUROQS,Kalaee:2021:QTURPRE,Monnai:2022:QTUR,Hasegawa:2023:BulkBoundaryBoundNC,Nishiyama:2024:OpenQuantumRURJPA,
Hasegawa:2024:ConcentrationIneqPRL,
Prech:2025:CoherenceQTUR} have garnered significant attention. In particular, the thermodynamic uncertainty relations within the framework of continuous measurement are closely related to those in classical stochastic processes. This direction presents future challenges.

\appendix

\section{Derivation of the main result [Eqs. \eqref{eq:main_result1} and \eqref{eq:main_result2}]\label{sec:derivation_main}}

Let us introduce the absolute value of the random variable $\Phi$, denoted by $|\Phi|$.
The probability distribution of $|\Phi|$ is given by
\begin{align}
    P(|\Phi|=\phi)=\begin{cases}
P(\Phi=\phi)+P(\Phi=-\phi) & \phi>0\\
P(\Phi=0) & \phi=0
\end{cases}
\label{eq:P_absX_def}
\end{align}
Therefore, the Shannon entropy of $|\Phi|$ is given by
\begin{align}
    &H[P(|\Phi|)]\nonumber\\&=-\sum_{\phi\ge0}P(|\Phi|=\phi)\ln P(|\Phi|=\phi)\nonumber\\&=-P(0)\ln P(0)-\sum_{\phi>0}[P(\phi)+P(-\phi)]\ln[P(\phi)+P(-\phi)]\nonumber\\&=-P(0)\ln P(0)-\sum_{\phi>0}P(\phi)\ln[P(\phi)+P(-\phi)]\nonumber\\&-\sum_{\phi<0}P(\phi)\ln[P(\phi)+P(-\phi)]\nonumber\\&=-\sum_{\phi}P(\phi)\ln[P(\phi)+P(-\phi)]+P(0)\ln2,
    \label{eq:H_Pabs_supp}
\end{align}
where we abbreviated $P(\phi) = P(\Phi = \phi)$.

Using the monotonicity of the Kullback-Leibler divergence [Eq.~\eqref{eq:KL_monotonicity}] and the time-reversal property of $\Phi$ [Eq.~\eqref{eq:time_reversal}], we have
\begin{align}
    \Sigma&\ge D[P(\Phi)\|P(-\Phi)]\nonumber\\&=-H[P(\Phi)]+C[P(\Phi),P(-\Phi)].
    \label{eq:Sigma_monotoniciy}
\end{align}
Here, the cross entropy is evaluated as $C[P(\Phi),P(-\Phi)]=-\sum_{\phi}P(\Phi=\phi)\ln P(\Phi=-\phi)$. 
Since the cross entropy term is no less than the entropy, it is non-negative, $C[P(\Phi),P(-\Phi)]\ge H[P(\Phi)]\ge0$. 
The important observation is that $C[P(\Phi),P(-\Phi)]$ is even bounded from below by a positive term. 
We calculate $H[P(|\Phi|)]-C[P(\Phi),P(-\Phi)]$ as follows:
\begin{align}
    &H[P(|\Phi|)]-C[P(\Phi),P(-\Phi)]\nonumber\\&=\sum_{\phi}P(\phi)\ln\frac{P(-\phi)/P(\phi)}{1+P(-\phi)/P(\phi)}+P(0)\ln2.
    \label{eq:H_minus_C_supp}
\end{align}
Let us consider the function
\begin{align}
    \mathfrak{f}(x)=\ln\frac{x}{1+x}.
    \label{eq:fx_def_appendix}
\end{align}
Since $\mathfrak{f}(x)$ is concave for $x>0$, by using the Jensen inequality, the following relation holds:
\begin{align}
    &\sum_{\phi}P(\phi)\ln\frac{P(-\phi)/P(\phi)}{1+P(-\phi)/P(\phi)}\nonumber\\&\le\ln\frac{\sum_{\phi}P(\phi)\left(P(-\phi)/P(\phi)\right)}{1+\sum_{\phi}P(\phi)\left(P(-\phi)/P(\phi)\right)}\nonumber\\&=-\ln2.
    \label{eq:C_Jensen_supp}
\end{align}
By substituting Eq.~\eqref{eq:C_Jensen_supp} into Eq.~\eqref{eq:H_minus_C_supp}, we obtain 
\begin{align}
    &H[P(|\Phi|)]+[1-P(\Phi=0)]\ln2\nonumber\\&\le C[P(\Phi),P(-\Phi)].
    \label{eq:cross_ineq_supp}
\end{align}
Substituting Eq.~\eqref{eq:cross_ineq_supp} into Eq.~\eqref{eq:Sigma_monotoniciy},
Equation~\eqref{eq:cross_ineq_supp} proves Eqs.~\eqref{eq:main_result1} and \eqref{eq:main_result2} in the main text. 

Next, we consider the continuous case. 
Basically, the derivation is the same as in the discrete case except that summation is replaced by integration.
Assume $P(\Phi)$ represents a probability density that is differentiable for all $\Phi$.
The Shannon entropy is defined by
\begin{align}
    H[P(\Phi)] = -\int_{-\infty}^{\infty} d\phi\; P(\phi) \ln P(\phi).
    \label{eq:Shannon_continuous}
\end{align}
For the continuous case, the probability density $P(|\Phi|)$ is defined by
\begin{align}
    P(|\Phi|=\phi)=P(\Phi=\phi)+P(\Phi=-\phi)\;\;(\phi\ge0).
    \label{eq:CV_P_absX_def}
\end{align}
We do not consider $P(|\Phi| = 0)$, because this set has measure zero for the smooth probability density. 
When $P(\Phi)$ includes the contribution of the delta function in $\Phi=0$, this is not the case. 
Following the same procedure as the discrete case, we obtain
\begin{align}
    H[P(|\Phi|)]+\ln2&\le C[P(\Phi),P(-\Phi)].
    \label{eq:CV_cross_ineq_supp}
\end{align}

\section{Proof of Eqs.~\eqref{eq:R_range} and \eqref{eq:R_range2}\label{sec:derivation_R_range}}

In this section, we prove Eqs.~\eqref{eq:R_range} and \eqref{eq:R_range2}.
Here, we show the relation for $P(\Phi = 0) \ge 0$.
The relation that we want to show is given by
\begin{align}
    0\le \Lambda[P(\Phi)]\le[1-P(\Phi=0)]\ln2.
    \label{eq:R_range_supp}
\end{align}
The first inequality part corresponds to $H[P(|\Phi|)] \le H[P(\Phi)]$. 
Specifically, $H[P(\Phi)]-H[P(|\Phi|)]$ is 
\begin{align}
&H[P(\Phi)]-H[P(|\Phi|)]\nonumber\\&=\sum_{{\phi>0}}[P(\phi)+P(-\phi)]\ln\left[P(\phi)+P(-\phi)\right]\nonumber\\&{+P(0)\ln P(0)}-\sum_{\phi}P(\phi)\ln P(\phi)\nonumber\\&=\sum_{\phi>0}\Bigl[[P(\phi)+P(-\phi)]\ln\left[P(\phi)+P(-\phi)\right]\nonumber\\&-P(\phi)\ln P(\phi)-P(-\phi)\ln P(-\phi)\Bigr]\ge0,
    \label{eq:H_minus_Habs}
\end{align}
where we used 
$(a+b)\ln(a+b)-a\ln a-b\ln b>0$ for $a>0$ and $b>0$ in the last line. 
$(a+b)\ln(a+b)-a\ln a-b\ln b = 0$ if and only if $a = 0$ or $b = 0$.
This indicates that Eq.~\eqref{eq:H_minus_Habs} saturates if and only if $P(\phi) = 0$ or $P(-\phi) = 0$ for all $\phi$ except $P(0)$. 

Next, we prove the second part of the inequality of Eq.~\eqref{eq:R_range_supp}, which can be done following the same approach as in Appendix~\ref{sec:derivation_main}. 
Using Eq.~\eqref{eq:H_Pabs_supp}, we have
\begin{align}
    &H[P(|\Phi|)]-H[P(\Phi)]\nonumber\\&=-\sum_{\phi}P(\phi)\ln\left(1+\frac{P(-\phi)}{P(\phi)}\right)+P(0)\ln2\nonumber\\&\ge-\ln2+P(0)\ln2,
    \label{eq:HPabs_HP_diff}
\end{align}
where we again used the Jensen inequality. 
Equation~\eqref{eq:HPabs_HP_diff} proves the second inequality part of Eq.~\eqref{eq:R_range_supp}. 
Equality in Eq.~\eqref{eq:HPabs_HP_diff} holds if and only if $P(-\phi)/P(\phi) = 1$. 

By adopting the formulation based on the Jensen-Shannon divergence [Eq.~\eqref{eq:rhs_JS}], the second part of the inequality in Eq.~\eqref{eq:R_range_supp} can be directly derived from Eq.~\eqref{eq:JS_div_bound}.
However, note that the first inequality part of Eq.~\eqref{eq:R_range_supp} cannot be obtained from Eq.~\eqref{eq:JS_div_bound}.

\section{Symmetry entropy and Jensen-Shannon divergence\label{sec:symmetry_entropy_Jensen_Shannon}}

We derive Eq.~\eqref{eq:rhs_JS}. Although we assume $P(\Phi=0)=0$ in Eq.~\eqref{eq:rhs_JS}, the derivation below also applies to the general case, including $P(\Phi=0)>0$.
The Jensen-Shannon divergence is defined by
\begin{align}
    \mathrm{JS}[P(X)\|P(Y)]&\equiv\frac{1}{2}D\left[P(X)\middle\|\frac{P(X)+P(Y)}{2}\right]\nonumber\\&+\frac{1}{2}D\left[P(Y)\middle\|\frac{P(X)+P(Y)}{2}\right],
    \label{eq:JS_div_def}
\end{align}
which satisfies
\begin{align}
    0 \le \mathrm{JS}[P(X)\|P(Y)] \le \ln 2.
    \label{eq:JS_div_bound}
\end{align}
Using the Jensen-Shannon divergence, we have 
\begin{align}
    &\mathrm{JS}[P(\Phi)\|P(-\Phi)]\nonumber\\&=-\sum_{\phi}\frac{P(\phi)+P(-\phi)}{2}\ln\frac{P(\phi)+P(-\phi)}{2}\nonumber\\&+\frac{1}{2}\sum_{\phi}P(\phi)\ln P(\phi)+\frac{1}{2}\sum_{\phi}P(-\phi)\ln P(-\phi)\nonumber\\&=\ln2-\sum_{\phi}P(\phi)\ln\left[P(\phi)+P(-\phi)\right]+\sum_{\phi}P(\phi)\ln P(\phi)\nonumber\\&=\ln2+H[P(|\Phi|)]-P(0)\ln2-H[P(\Phi)]\nonumber\\&=\ln2-P(0)\ln2-\Lambda[P(\Phi)],
    \label{eq:J_ln2_Lambda}
\end{align}
where we used Eq.~\eqref{eq:H_Pabs_supp}. 
Equation~\eqref{eq:J_ln2_Lambda} is Eq.~\eqref{eq:rhs_JS} in the main text.

\section{Equality condition\label{sec:equality_condition}}

We consider the equality condition of the thermodynamic entropic uncertainty relation [Eqs.~\eqref{eq:main_result1} and \eqref{eq:main_result2}]. 
The derivation of the thermodynamic entropic uncertainty relation employs two key inequalities:
\begin{enumerate}
  \item Monotonicity of the Kullback-Leibler divergence under coarse graining.
  \item The Jensen inequality applied to a strictly concave function.
\end{enumerate}
To saturate the thermodynamic entropic uncertainty relation, both inequalities must be saturated simultaneously.
The distribution $P(\phi)$ is defined by
\begin{align}
    P(\phi)=\sum_{\Gamma}\delta\bigl(\Phi(\Gamma) - \phi\bigr)\mathcal{P}(\Gamma).
    \label{eq:Pphi_def_supp}
\end{align}
Then the monotonicity states
\begin{align}
    D\bigl[\mathcal{P}(\Gamma)\Vert\mathcal{P}(\Gamma^{\dagger})\bigr]\ge D\bigl[P(\phi)\Vert P(-\phi)\bigr].
\label{eq:monotonicity}
\end{align}
Equality holds if and only if the log-ratio depends on $\Gamma$ solely through $\Phi(\Gamma)$, i.e., there exists a function $h$ such that
\begin{equation}
  \frac{\mathcal{P}(\Gamma)}{\mathcal{P}(\Gamma^{\dagger})}
  \;=\;
  h\!\bigl(\Phi(\Gamma)\bigr).
\label{eq:equality_condition1}
\end{equation}
There are two scenarios in which Eq.~\eqref{eq:equality_condition1} is satisfied:
\begin{description}
  \item[Case 1 (Equilibrium)] $\mathcal{P}(\Gamma)=\mathcal{P}(\Gamma^{\dagger})$ for all $\Gamma$.
  \item[Case 2 (Bijectivity)] $\Phi(\Gamma)$ is a bijective function.
\end{description}

Next, we consider the Jensen inequality part. 
We applied the Jensen inequality to the strictly concave function 
$\mathfrak{f}(x)$ defined in Eq.~\eqref{eq:fx_def_appendix}. 
The equality of the Jensen inequality is saturated if and only if the following condition is satisfied:
\begin{equation}
  \frac{P(\phi)}{P(-\phi)} \;=\;\text{constant}.
\label{eq:equality_condition2pre}
\end{equation}
Since Eq.~\eqref{eq:equality_condition2pre} should hold for all $\phi$, 
Eq.~\eqref{eq:equality_condition2pre} is equivalently stated as
\begin{align}
    P(\phi)=P(-\phi).
    \label{eq:equality_condition2}
\end{align}
To saturate the thermodynamic entropic uncertainty relation, we require both \eqref{eq:equality_condition1} and \eqref{eq:equality_condition2}.  We now show that this forces equilibrium:
\begin{itemize}
  \item In Case 1, $\mathcal{P}(\Gamma)=\mathcal{P}(\Gamma^{\dagger})$ already defines the equilibrium.  Moreover, equilibrium implies $P(\phi)=P(-\phi)$, so Eq.~\eqref{eq:equality_condition2} is automatically satisfied.
  \item In Case 2, $\Phi$ is bijective, so Eq.~\eqref{eq:equality_condition1} allows any $\mathcal{P}(\Gamma)$ expressed via $\Phi(\Gamma)$.  But Eq.~\eqref{eq:equality_condition2} then demands $\mathcal{P}(\Gamma)/\mathcal{P}(\Gamma^\dagger)$ be constant, again forcing $\mathcal{P}(\Gamma)=\mathcal{P}(\Gamma^{\dagger})$.
\end{itemize}

In summary, the thermodynamic entropic uncertainty relation [Eqs.~\eqref{eq:main_result1} and \eqref{eq:main_result2}] is saturated if and only if
the system is in equilibrium.

\section{Diffusion decision model\label{sec:DDM}}

\subsection{Model\label{sec:model}}

In the main text, we consider the diffusion decision model,
where the evidence accumulation process is modeled by
Eq.~\eqref{eq:dzdt_def}. 
Let $P(x;t)$ be the probability density of $x$ at time $t$. 
The corresponding Fokker-Planck equation is 
\begin{align}
    \frac{\partial P(x; t)}{\partial t}=-\frac{\partial}{\partial x}[f P(x; t)]+A \frac{\partial^2 P(x; t)}{\partial x^2}.
    \label{eq:FPE_supp}
\end{align}
The probability current of Eq.~\eqref{eq:FPE_supp} is 
\begin{align}
    J(x;t)\equiv fP(x;t)-A\frac{\partial P(x;t)}{\partial x}.
    \label{eq:Jxt_def}
\end{align}
A trajectory $\Gamma$ corresponds to a realization of the Langevin dynamics within the time interval $[0,\tau]$. 
In the main text, we define the observable $\Phi(\Gamma)$ for a decision-making process. 
The probability density of $y(t)$ is given by the Gaussian distribution \cite{Risken:1989:FPEBook}:
\begin{align}
    P(y;t)=\frac{1}{\sqrt{4\pi At}}\exp\left[-\frac{\left(y-ft\right)^{2}}{4At}\right].
    \label{eq:Pxt_def}
\end{align}

\subsection{Discrete case\label{sec:discrete_case}}

In the discrete model, the observable is defined in Eq.~\eqref{eq:Phi_def_main}. 
From Eq.~\eqref{eq:Pxt_def}, 
the observable probability distribution at time $\tau$ is
\begin{align}
P(\Phi=+1)&=\int_{\theta}^{\infty}P(y;t)dy=\frac{1}{2}\mathrm{erf}\left(\frac{ft-\theta}{2\sqrt{At}}\right)+\frac{1}{2},\label{eq:obs_plus_def}\\P(\Phi=0)&=\int_{-\theta}^{\theta}P(y;t)dy=-\frac{1}{2}\mathrm{erf}\left(\frac{ft-\theta}{2\sqrt{At}}\right)\nonumber\\&+\frac{1}{2}\mathrm{erf}\left(\frac{ft+\theta}{2\sqrt{At}}\right),\label{eq:obs_zero_def}\\P(\Phi=-1)&=\int_{-\infty}^{-\theta}P(y;t)dy=\frac{1}{2}-\frac{1}{2}\mathrm{erf}\left(\frac{ft+\theta}{2\sqrt{At}}\right).\label{eq:obs_minus_def}
\end{align}
The probabilities in Eqs.~\eqref{eq:obs_plus_def}--\eqref{eq:obs_minus_def} are used to calculate $\Lambda[P(\Phi)]$.

\subsection{Continuous case\label{sec:continuous_case}}

The symmetry entropy is given by $\Lambda[P(\Phi)] = H[P(\Phi)]-H[P(|\Phi|)]$ [Eq.~\eqref{eq:R_def}].
For the observable given by Eq.~\eqref{eq:Phi_continuous}, each of the terms in $\Lambda[P(\Phi)]$ is
\begin{align}
    H[P(\Phi)]&=-\int_{-\infty}^{\infty}dy\,P(y;\tau)\ln P(y;\tau),\label{eq:HPPhi_supp}\\H[P(|\Phi|)]&=-\int_{0}^{\infty}dy\,\Bigl[\left[P(y;\tau)+P(-y;\tau)\right]\nonumber\\&\times\ln\left[P(y;\tau)+P(-y;\tau)\right]\Bigr],\label{eq:HPPhi_abs_supp}
\end{align}
where $P(y;t)$ is Eq.~\eqref{eq:Pxt_def}.

\begin{acknowledgments}

This work was supported by JSPS KAKENHI Grant Number JP23K24915.

\end{acknowledgments}

\subsection*{DATA AVAILABILITY}

The data that support the findings of this article are openly available \cite{Data:2025:TEUR}.


\begin{thebibliography}{50}%
\makeatletter
\providecommand \@ifxundefined [1]{%
 \@ifx{#1\undefined}
}%
\providecommand \@ifnum [1]{%
 \ifnum #1\expandafter \@firstoftwo
 \else \expandafter \@secondoftwo
 \fi
}%
\providecommand \@ifx [1]{%
 \ifx #1\expandafter \@firstoftwo
 \else \expandafter \@secondoftwo
 \fi
}%
\providecommand \natexlab [1]{#1}%
\providecommand \enquote  [1]{``#1''}%
\providecommand \bibnamefont  [1]{#1}%
\providecommand \bibfnamefont [1]{#1}%
\providecommand \citenamefont [1]{#1}%
\providecommand \href@noop [0]{\@secondoftwo}%
\providecommand \href [0]{\begingroup \@sanitize@url \@href}%
\providecommand \@href[1]{\@@startlink{#1}\@@href}%
\providecommand \@@href[1]{\endgroup#1\@@endlink}%
\providecommand \@sanitize@url [0]{\catcode `\\12\catcode `\$12\catcode `\&12\catcode `\#12\catcode `\^12\catcode `\_12\catcode `\%12\relax}%
\providecommand \@@startlink[1]{}%
\providecommand \@@endlink[0]{}%
\providecommand \url  [0]{\begingroup\@sanitize@url \@url }%
\providecommand \@url [1]{\endgroup\@href {#1}{\urlprefix }}%
\providecommand \urlprefix  [0]{URL }%
\providecommand \Eprint [0]{\href }%
\providecommand \doibase [0]{https://doi.org/}%
\providecommand \selectlanguage [0]{\@gobble}%
\providecommand \bibinfo  [0]{\@secondoftwo}%
\providecommand \bibfield  [0]{\@secondoftwo}%
\providecommand \translation [1]{[#1]}%
\providecommand \BibitemOpen [0]{}%
\providecommand \bibitemStop [0]{}%
\providecommand \bibitemNoStop [0]{.\EOS\space}%
\providecommand \EOS [0]{\spacefactor3000\relax}%
\providecommand \BibitemShut  [1]{\csname bibitem#1\endcsname}%
\let\auto@bib@innerbib\@empty
\bibitem [{\citenamefont {Barato}\ and\ \citenamefont {Seifert}(2015)}]{Barato:2015:UncRel}%
  \BibitemOpen
  \bibfield  {author} {\bibinfo {author} {\bibfnamefont {A.~C.}\ \bibnamefont {Barato}}\ and\ \bibinfo {author} {\bibfnamefont {U.}~\bibnamefont {Seifert}},\ }\bibfield  {title} {\bibinfo {title} {Thermodynamic uncertainty relation for biomolecular processes},\ }\href {https://doi.org/10.1103/PhysRevLett.114.158101} {\bibfield  {journal} {\bibinfo  {journal} {Phys. Rev. Lett.}\ }\textbf {\bibinfo {volume} {114}},\ \bibinfo {pages} {158101} (\bibinfo {year} {2015})}\BibitemShut {NoStop}%
\bibitem [{\citenamefont {Gingrich}\ \emph {et~al.}(2016)\citenamefont {Gingrich}, \citenamefont {Horowitz}, \citenamefont {Perunov},\ and\ \citenamefont {England}}]{Gingrich:2016:TUP}%
  \BibitemOpen
  \bibfield  {author} {\bibinfo {author} {\bibfnamefont {T.~R.}\ \bibnamefont {Gingrich}}, \bibinfo {author} {\bibfnamefont {J.~M.}\ \bibnamefont {Horowitz}}, \bibinfo {author} {\bibfnamefont {N.}~\bibnamefont {Perunov}},\ and\ \bibinfo {author} {\bibfnamefont {J.~L.}\ \bibnamefont {England}},\ }\bibfield  {title} {\bibinfo {title} {Dissipation bounds all steady-state current fluctuations},\ }\href {https://doi.org/10.1103/PhysRevLett.116.120601} {\bibfield  {journal} {\bibinfo  {journal} {Phys. Rev. Lett.}\ }\textbf {\bibinfo {volume} {116}},\ \bibinfo {pages} {120601} (\bibinfo {year} {2016})}\BibitemShut {NoStop}%
\bibitem [{\citenamefont {Garrahan}(2017)}]{Garrahan:2017:TUR}%
  \BibitemOpen
  \bibfield  {author} {\bibinfo {author} {\bibfnamefont {J.~P.}\ \bibnamefont {Garrahan}},\ }\bibfield  {title} {\bibinfo {title} {Simple bounds on fluctuations and uncertainty relations for first-passage times of counting observables},\ }\href {https://doi.org/10.1103/PhysRevE.95.032134} {\bibfield  {journal} {\bibinfo  {journal} {Phys. Rev. E}\ }\textbf {\bibinfo {volume} {95}},\ \bibinfo {pages} {032134} (\bibinfo {year} {2017})}\BibitemShut {NoStop}%
\bibitem [{\citenamefont {Dechant}\ and\ \citenamefont {Sasa}(2018)}]{Dechant:2018:TUR}%
  \BibitemOpen
  \bibfield  {author} {\bibinfo {author} {\bibfnamefont {A.}~\bibnamefont {Dechant}}\ and\ \bibinfo {author} {\bibfnamefont {S.-i.}\ \bibnamefont {Sasa}},\ }\bibfield  {title} {\bibinfo {title} {Current fluctuations and transport efficiency for general {Langevin} systems},\ }\href {https://doi.org/10.1088/1742-5468/aac91a} {\bibfield  {journal} {\bibinfo  {journal} {J. Stat. Mech: Theory Exp.}\ }\textbf {\bibinfo {volume} {2018}},\ \bibinfo {pages} {063209} (\bibinfo {year} {2018})}\BibitemShut {NoStop}%
\bibitem [{\citenamefont {{Di Terlizzi}}\ and\ \citenamefont {Baiesi}(2019)}]{Terlizzi:2019:KUR}%
  \BibitemOpen
  \bibfield  {author} {\bibinfo {author} {\bibfnamefont {I.}~\bibnamefont {{Di Terlizzi}}}\ and\ \bibinfo {author} {\bibfnamefont {M.}~\bibnamefont {Baiesi}},\ }\bibfield  {title} {\bibinfo {title} {Kinetic uncertainty relation},\ }\href {https://doi.org/10.1088/1751-8121/aaee34} {\bibfield  {journal} {\bibinfo  {journal} {J. Phys. A: Math. Theor.}\ }\textbf {\bibinfo {volume} {52}},\ \bibinfo {pages} {02LT03} (\bibinfo {year} {2019})}\BibitemShut {NoStop}%
\bibitem [{\citenamefont {Hasegawa}\ and\ \citenamefont {Van~Vu}(2019{\natexlab{a}})}]{Hasegawa:2019:CRI}%
  \BibitemOpen
  \bibfield  {author} {\bibinfo {author} {\bibfnamefont {Y.}~\bibnamefont {Hasegawa}}\ and\ \bibinfo {author} {\bibfnamefont {T.}~\bibnamefont {Van~Vu}},\ }\bibfield  {title} {\bibinfo {title} {Uncertainty relations in stochastic processes: An information inequality approach},\ }\href {https://doi.org/10.1103/PhysRevE.99.062126} {\bibfield  {journal} {\bibinfo  {journal} {Phys. Rev. E}\ }\textbf {\bibinfo {volume} {99}},\ \bibinfo {pages} {062126} (\bibinfo {year} {2019}{\natexlab{a}})}\BibitemShut {NoStop}%
\bibitem [{\citenamefont {Hasegawa}\ and\ \citenamefont {Van~Vu}(2019{\natexlab{b}})}]{Hasegawa:2019:FTUR}%
  \BibitemOpen
  \bibfield  {author} {\bibinfo {author} {\bibfnamefont {Y.}~\bibnamefont {Hasegawa}}\ and\ \bibinfo {author} {\bibfnamefont {T.}~\bibnamefont {Van~Vu}},\ }\bibfield  {title} {\bibinfo {title} {Fluctuation theorem uncertainty relation},\ }\href {https://doi.org/10.1103/PhysRevLett.123.110602} {\bibfield  {journal} {\bibinfo  {journal} {Phys. Rev. Lett.}\ }\textbf {\bibinfo {volume} {123}},\ \bibinfo {pages} {110602} (\bibinfo {year} {2019}{\natexlab{b}})}\BibitemShut {NoStop}%
\bibitem [{\citenamefont {Dechant}\ and\ \citenamefont {Sasa}(2020)}]{Dechant:2020:FRIPNAS}%
  \BibitemOpen
  \bibfield  {author} {\bibinfo {author} {\bibfnamefont {A.}~\bibnamefont {Dechant}}\ and\ \bibinfo {author} {\bibfnamefont {S.-i.}\ \bibnamefont {Sasa}},\ }\bibfield  {title} {\bibinfo {title} {Fluctuation--response inequality out of equilibrium},\ }\href {https://doi.org/10.1073/pnas.1918386117} {\bibfield  {journal} {\bibinfo  {journal} {Proc. Natl. Acad. Sci. U.S.A.}\ }\textbf {\bibinfo {volume} {117}},\ \bibinfo {pages} {6430} (\bibinfo {year} {2020})}\BibitemShut {NoStop}%
\bibitem [{\citenamefont {Vo}\ \emph {et~al.}(2020)\citenamefont {Vo}, \citenamefont {Van~Vu},\ and\ \citenamefont {Hasegawa}}]{Vo:2020:TURCSLPRE}%
  \BibitemOpen
  \bibfield  {author} {\bibinfo {author} {\bibfnamefont {V.~T.}\ \bibnamefont {Vo}}, \bibinfo {author} {\bibfnamefont {T.}~\bibnamefont {Van~Vu}},\ and\ \bibinfo {author} {\bibfnamefont {Y.}~\bibnamefont {Hasegawa}},\ }\bibfield  {title} {\bibinfo {title} {Unified approach to classical speed limit and thermodynamic uncertainty relation},\ }\href {https://doi.org/10.1103/PhysRevE.102.062132} {\bibfield  {journal} {\bibinfo  {journal} {Phys. Rev. E}\ }\textbf {\bibinfo {volume} {102}},\ \bibinfo {pages} {062132} (\bibinfo {year} {2020})}\BibitemShut {NoStop}%
\bibitem [{\citenamefont {Koyuk}\ and\ \citenamefont {Seifert}(2020)}]{Koyuk:2020:TUR}%
  \BibitemOpen
  \bibfield  {author} {\bibinfo {author} {\bibfnamefont {T.}~\bibnamefont {Koyuk}}\ and\ \bibinfo {author} {\bibfnamefont {U.}~\bibnamefont {Seifert}},\ }\bibfield  {title} {\bibinfo {title} {Thermodynamic uncertainty relation for time-dependent driving},\ }\href {https://doi.org/10.1103/PhysRevLett.125.260604} {\bibfield  {journal} {\bibinfo  {journal} {Phys. Rev. Lett.}\ }\textbf {\bibinfo {volume} {125}},\ \bibinfo {pages} {260604} (\bibinfo {year} {2020})}\BibitemShut {NoStop}%
\bibitem [{\citenamefont {Heisenberg}(1927)}]{Heisenberg:1927:UR}%
  \BibitemOpen
  \bibfield  {author} {\bibinfo {author} {\bibfnamefont {W.}~\bibnamefont {Heisenberg}},\ }\bibfield  {title} {\bibinfo {title} {{\"U}ber den anschaulichen inhalt der quantentheoretischen kinematik und mechanik},\ }\href {https://doi.org/10.1007/BF01397280} {\bibfield  {journal} {\bibinfo  {journal} {Z. Phys.}\ }\textbf {\bibinfo {volume} {43}},\ \bibinfo {pages} {172} (\bibinfo {year} {1927})}\BibitemShut {NoStop}%
\bibitem [{\citenamefont {Robertson}(1929)}]{Robertson:1929:UncRel}%
  \BibitemOpen
  \bibfield  {author} {\bibinfo {author} {\bibfnamefont {H.~P.}\ \bibnamefont {Robertson}},\ }\bibfield  {title} {\bibinfo {title} {The uncertainty principle},\ }\href {https://doi.org/10.1103/PhysRev.34.163} {\bibfield  {journal} {\bibinfo  {journal} {Phys. Rev.}\ }\textbf {\bibinfo {volume} {34}},\ \bibinfo {pages} {163} (\bibinfo {year} {1929})}\BibitemShut {NoStop}%
\bibitem [{\citenamefont {Hirschman}(1957)}]{Hirschman:1957:EntropyUncertainty}%
  \BibitemOpen
  \bibfield  {author} {\bibinfo {author} {\bibfnamefont {I.~I.}\ \bibnamefont {Hirschman}},\ }\bibfield  {title} {\bibinfo {title} {A note on entropy},\ }\href {https://doi.org/10.2307/2372390} {\bibfield  {journal} {\bibinfo  {journal} {Amer. J. Math.}\ }\textbf {\bibinfo {volume} {79}},\ \bibinfo {pages} {152} (\bibinfo {year} {1957})}\BibitemShut {NoStop}%
\bibitem [{\citenamefont {Bia{\l}ynicki-Birula}\ and\ \citenamefont {Mycielski}(1975)}]{BialynickiBirula:1975:EUR}%
  \BibitemOpen
  \bibfield  {author} {\bibinfo {author} {\bibfnamefont {I.}~\bibnamefont {Bia{\l}ynicki-Birula}}\ and\ \bibinfo {author} {\bibfnamefont {J.}~\bibnamefont {Mycielski}},\ }\bibfield  {title} {\bibinfo {title} {Uncertainty relations for information entropy in wave mechanics},\ }\href {https://doi.org/10.1007/BF01608825} {\bibfield  {journal} {\bibinfo  {journal} {Comm. Math. Phys.}\ }\textbf {\bibinfo {volume} {44}},\ \bibinfo {pages} {129} (\bibinfo {year} {1975})}\BibitemShut {NoStop}%
\bibitem [{\citenamefont {Maassen}\ and\ \citenamefont {Uffink}(1988)}]{Maassen:1988:Entropic}%
  \BibitemOpen
  \bibfield  {author} {\bibinfo {author} {\bibfnamefont {H.}~\bibnamefont {Maassen}}\ and\ \bibinfo {author} {\bibfnamefont {J.~B.~M.}\ \bibnamefont {Uffink}},\ }\bibfield  {title} {\bibinfo {title} {Generalized entropic uncertainty relations},\ }\href {https://doi.org/10.1103/PhysRevLett.60.1103} {\bibfield  {journal} {\bibinfo  {journal} {Phys. Rev. Lett.}\ }\textbf {\bibinfo {volume} {60}},\ \bibinfo {pages} {1103} (\bibinfo {year} {1988})}\BibitemShut {NoStop}%
\bibitem [{\citenamefont {Coles}\ \emph {et~al.}(2017)\citenamefont {Coles}, \citenamefont {Berta}, \citenamefont {Tomamichel},\ and\ \citenamefont {Wehner}}]{Coles:2017:EntropicUR}%
  \BibitemOpen
  \bibfield  {author} {\bibinfo {author} {\bibfnamefont {P.~J.}\ \bibnamefont {Coles}}, \bibinfo {author} {\bibfnamefont {M.}~\bibnamefont {Berta}}, \bibinfo {author} {\bibfnamefont {M.}~\bibnamefont {Tomamichel}},\ and\ \bibinfo {author} {\bibfnamefont {S.}~\bibnamefont {Wehner}},\ }\bibfield  {title} {\bibinfo {title} {Entropic uncertainty relations and their applications},\ }\href {https://doi.org/10.1103/RevModPhys.89.015002} {\bibfield  {journal} {\bibinfo  {journal} {Rev. Mod. Phys.}\ }\textbf {\bibinfo {volume} {89}},\ \bibinfo {pages} {015002} (\bibinfo {year} {2017})}\BibitemShut {NoStop}%
\bibitem [{\citenamefont {Berta}\ \emph {et~al.}(2010)\citenamefont {Berta}, \citenamefont {Christandl}, \citenamefont {Colbeck}, \citenamefont {Renes},\ and\ \citenamefont {Renner}}]{Berta:2010:MemoryEUR}%
  \BibitemOpen
  \bibfield  {author} {\bibinfo {author} {\bibfnamefont {M.}~\bibnamefont {Berta}}, \bibinfo {author} {\bibfnamefont {M.}~\bibnamefont {Christandl}}, \bibinfo {author} {\bibfnamefont {R.}~\bibnamefont {Colbeck}}, \bibinfo {author} {\bibfnamefont {J.~M.}\ \bibnamefont {Renes}},\ and\ \bibinfo {author} {\bibfnamefont {R.}~\bibnamefont {Renner}},\ }\bibfield  {title} {\bibinfo {title} {The uncertainty principle in the presence of quantum memory},\ }\href {https://doi.org/10.1038/nphys1734} {\bibfield  {journal} {\bibinfo  {journal} {Nat. Phys.}\ }\textbf {\bibinfo {volume} {6}},\ \bibinfo {pages} {659} (\bibinfo {year} {2010})}\BibitemShut {NoStop}%
\bibitem [{\citenamefont {Shiraishi}\ \emph {et~al.}(2018)\citenamefont {Shiraishi}, \citenamefont {Funo},\ and\ \citenamefont {Saito}}]{Shiraishi:2018:SpeedLimit}%
  \BibitemOpen
  \bibfield  {author} {\bibinfo {author} {\bibfnamefont {N.}~\bibnamefont {Shiraishi}}, \bibinfo {author} {\bibfnamefont {K.}~\bibnamefont {Funo}},\ and\ \bibinfo {author} {\bibfnamefont {K.}~\bibnamefont {Saito}},\ }\bibfield  {title} {\bibinfo {title} {Speed limit for classical stochastic processes},\ }\href {https://link.aps.org/doi/10.1103/PhysRevLett.121.070601} {\bibfield  {journal} {\bibinfo  {journal} {Phys. Rev. Lett.}\ }\textbf {\bibinfo {volume} {121}},\ \bibinfo {pages} {070601} (\bibinfo {year} {2018})}\BibitemShut {NoStop}%
\bibitem [{\citenamefont {Ito}\ and\ \citenamefont {Dechant}(2020)}]{Ito:2020:TimeTURPRX}%
  \BibitemOpen
  \bibfield  {author} {\bibinfo {author} {\bibfnamefont {S.}~\bibnamefont {Ito}}\ and\ \bibinfo {author} {\bibfnamefont {A.}~\bibnamefont {Dechant}},\ }\bibfield  {title} {\bibinfo {title} {Stochastic time evolution, information geometry, and the {Cram\'er}-{Rao} bound},\ }\href {https://doi.org/10.1103/PhysRevX.10.021056} {\bibfield  {journal} {\bibinfo  {journal} {Phys. Rev. X}\ }\textbf {\bibinfo {volume} {10}},\ \bibinfo {pages} {021056} (\bibinfo {year} {2020})}\BibitemShut {NoStop}%
\bibitem [{\citenamefont {Ito}(2018)}]{Ito:2018:InfoGeo}%
  \BibitemOpen
  \bibfield  {author} {\bibinfo {author} {\bibfnamefont {S.}~\bibnamefont {Ito}},\ }\bibfield  {title} {\bibinfo {title} {Stochastic thermodynamic interpretation of information geometry},\ }\href {https://doi.org/10.1103/PhysRevLett.121.030605} {\bibfield  {journal} {\bibinfo  {journal} {Phys. Rev. Lett.}\ }\textbf {\bibinfo {volume} {121}},\ \bibinfo {pages} {030605} (\bibinfo {year} {2018})}\BibitemShut {NoStop}%
\bibitem [{\citenamefont {Van~Vu}\ and\ \citenamefont {Hasegawa}(2021)}]{Vu:2021:GeomBound}%
  \BibitemOpen
  \bibfield  {author} {\bibinfo {author} {\bibfnamefont {T.}~\bibnamefont {Van~Vu}}\ and\ \bibinfo {author} {\bibfnamefont {Y.}~\bibnamefont {Hasegawa}},\ }\bibfield  {title} {\bibinfo {title} {Geometrical bounds of the irreversibility in {Markovian} systems},\ }\href {https://doi.org/10.1103/PhysRevLett.126.010601} {\bibfield  {journal} {\bibinfo  {journal} {Phys. Rev. Lett.}\ }\textbf {\bibinfo {volume} {126}},\ \bibinfo {pages} {010601} (\bibinfo {year} {2021})}\BibitemShut {NoStop}%
\bibitem [{\citenamefont {Dechant}\ and\ \citenamefont {Sakurai}(2019)}]{Dechant:2019:Wasserstein}%
  \BibitemOpen
  \bibfield  {author} {\bibinfo {author} {\bibfnamefont {A.}~\bibnamefont {Dechant}}\ and\ \bibinfo {author} {\bibfnamefont {Y.}~\bibnamefont {Sakurai}},\ }\bibfield  {title} {\bibinfo {title} {Thermodynamic interpretation of {Wasserstein} distance},\ }\href {https://arxiv.org/abs/1912.08405} {\bibfield  {journal} {\bibinfo  {journal} {arXiv:1912.08405}\ } (\bibinfo {year} {2019})}\BibitemShut {NoStop}%
\bibitem [{\citenamefont {Van~Vu}\ and\ \citenamefont {Saito}(2023)}]{Vu:2022:OptimalTransportPRX}%
  \BibitemOpen
  \bibfield  {author} {\bibinfo {author} {\bibfnamefont {T.}~\bibnamefont {Van~Vu}}\ and\ \bibinfo {author} {\bibfnamefont {K.}~\bibnamefont {Saito}},\ }\bibfield  {title} {\bibinfo {title} {Thermodynamic unification of optimal transport: Thermodynamic uncertainty relation, minimum dissipation, and thermodynamic speed limits},\ }\href {https://link.aps.org/doi/10.1103/PhysRevX.13.011013} {\bibfield  {journal} {\bibinfo  {journal} {Phys. Rev. X}\ }\textbf {\bibinfo {volume} {13}},\ \bibinfo {pages} {011013} (\bibinfo {year} {2023})}\BibitemShut {NoStop}%
\bibitem [{\citenamefont {Crooks}(1999)}]{Crooks:1999:CFT}%
  \BibitemOpen
  \bibfield  {author} {\bibinfo {author} {\bibfnamefont {G.~E.}\ \bibnamefont {Crooks}},\ }\bibfield  {title} {\bibinfo {title} {Entropy production fluctuation theorem and the nonequilibrium work relation for free energy differences},\ }\href {https://doi.org/10.1103/PhysRevE.60.2721} {\bibfield  {journal} {\bibinfo  {journal} {Phys. Rev. E}\ }\textbf {\bibinfo {volume} {60}},\ \bibinfo {pages} {2721} (\bibinfo {year} {1999})}\BibitemShut {NoStop}%
\bibitem [{\citenamefont {Seifert}(2005)}]{Seifert:2005:FT}%
  \BibitemOpen
  \bibfield  {author} {\bibinfo {author} {\bibfnamefont {U.}~\bibnamefont {Seifert}},\ }\bibfield  {title} {\bibinfo {title} {Entropy production along a stochastic trajectory and an integral fluctuation theorem},\ }\href {https://doi.org/10.1103/PhysRevLett.95.040602} {\bibfield  {journal} {\bibinfo  {journal} {Phys. Rev. Lett.}\ }\textbf {\bibinfo {volume} {95}},\ \bibinfo {pages} {040602} (\bibinfo {year} {2005})}\BibitemShut {NoStop}%
\bibitem [{Sup()}]{Supp:2025:TEUR}%
  \BibitemOpen
  \href@noop {} {}\bibinfo {note} {See Supplemental Material for details of calculations, which includes Ref.~\cite{Boyd:2004:ConvexOptimBook}.}\BibitemShut {Stop}%
\bibitem [{\citenamefont {Boyd}\ and\ \citenamefont {Vandenberghe}(2004)}]{Boyd:2004:ConvexOptimBook}%
  \BibitemOpen
  \bibfield  {author} {\bibinfo {author} {\bibfnamefont {S.}~\bibnamefont {Boyd}}\ and\ \bibinfo {author} {\bibfnamefont {L.}~\bibnamefont {Vandenberghe}},\ }\href {https://doi.org/10.1017/CBO9780511804441} {\emph {\bibinfo {title} {Convex Optimization}}}\ (\bibinfo  {publisher} {Cambridge University Press},\ \bibinfo {year} {2004})\BibitemShut {NoStop}%
\bibitem [{\citenamefont {Landauer}(1961)}]{Landauer:1961:LP}%
  \BibitemOpen
  \bibfield  {author} {\bibinfo {author} {\bibfnamefont {R.}~\bibnamefont {Landauer}},\ }\bibfield  {title} {\bibinfo {title} {Irreversibility and heat generation in the computing process},\ }\href {https://doi.org/10.1147/rd.53.0183} {\bibfield  {journal} {\bibinfo  {journal} {IBM J. Res. Dev.}\ }\textbf {\bibinfo {volume} {5}},\ \bibinfo {pages} {183} (\bibinfo {year} {1961})}\BibitemShut {NoStop}%
\bibitem [{\citenamefont {Reeb}\ and\ \citenamefont {Wolf}(2014)}]{Reed:2014:Landauer}%
  \BibitemOpen
  \bibfield  {author} {\bibinfo {author} {\bibfnamefont {D.}~\bibnamefont {Reeb}}\ and\ \bibinfo {author} {\bibfnamefont {M.~M.}\ \bibnamefont {Wolf}},\ }\bibfield  {title} {\bibinfo {title} {An improved {Landauer} principle with finite-size corrections},\ }\href {https://doi.org/10.1088/1367-2630/16/10/103011} {\bibfield  {journal} {\bibinfo  {journal} {New. J. Phys.}\ }\textbf {\bibinfo {volume} {16}},\ \bibinfo {pages} {103011} (\bibinfo {year} {2014})}\BibitemShut {NoStop}%
\bibitem [{\citenamefont {Ratcliff}\ and\ \citenamefont {McKoon}(2008)}]{Ratcliff:2008:DDM}%
  \BibitemOpen
  \bibfield  {author} {\bibinfo {author} {\bibfnamefont {R.}~\bibnamefont {Ratcliff}}\ and\ \bibinfo {author} {\bibfnamefont {G.}~\bibnamefont {McKoon}},\ }\bibfield  {title} {\bibinfo {title} {The diffusion decision model: Theory and data for two-choice decision tasks},\ }\href {https://doi.org/10.1162/neco.2008.12-06-420} {\bibfield  {journal} {\bibinfo  {journal} {Neural Comput}\ }\textbf {\bibinfo {volume} {20}},\ \bibinfo {pages} {873} (\bibinfo {year} {2008})}\BibitemShut {NoStop}%
\bibitem [{\citenamefont {Ratcliff}\ \emph {et~al.}(2016)\citenamefont {Ratcliff}, \citenamefont {Smith}, \citenamefont {Brown},\ and\ \citenamefont {McKoon}}]{Ratcliff:2016:DDM}%
  \BibitemOpen
  \bibfield  {author} {\bibinfo {author} {\bibfnamefont {R.}~\bibnamefont {Ratcliff}}, \bibinfo {author} {\bibfnamefont {P.~L.}\ \bibnamefont {Smith}}, \bibinfo {author} {\bibfnamefont {S.~D.}\ \bibnamefont {Brown}},\ and\ \bibinfo {author} {\bibfnamefont {G.}~\bibnamefont {McKoon}},\ }\bibfield  {title} {\bibinfo {title} {Diffusion decision model: Current issues and history},\ }\href {https://doi.org/10.1016/j.tics.2016.01.007} {\bibfield  {journal} {\bibinfo  {journal} {Trends Cogn. Sci.}\ }\textbf {\bibinfo {volume} {20}},\ \bibinfo {pages} {260} (\bibinfo {year} {2016})}\BibitemShut {NoStop}%
\bibitem [{\citenamefont {Myers}\ \emph {et~al.}(2022)\citenamefont {Myers}, \citenamefont {Interian},\ and\ \citenamefont {Moustafa}}]{Myers:2022:DDM}%
  \BibitemOpen
  \bibfield  {author} {\bibinfo {author} {\bibfnamefont {C.~E.}\ \bibnamefont {Myers}}, \bibinfo {author} {\bibfnamefont {A.}~\bibnamefont {Interian}},\ and\ \bibinfo {author} {\bibfnamefont {A.~A.}\ \bibnamefont {Moustafa}},\ }\bibfield  {title} {\bibinfo {title} {A practical introduction to using the drift diffusion model of decision-making in cognitive psychology, neuroscience, and health sciences},\ }\href {https://doi.org/10.3389/fpsyg.2022.1039172} {\bibfield  {journal} {\bibinfo  {journal} {Front. Psychol.}\ }\textbf {\bibinfo {volume} {13}} (\bibinfo {year} {2022})}\BibitemShut {NoStop}%
\bibitem [{\citenamefont {Stone}(1960)}]{Stone:1960:ChoiceReaction}%
  \BibitemOpen
  \bibfield  {author} {\bibinfo {author} {\bibfnamefont {M.}~\bibnamefont {Stone}},\ }\bibfield  {title} {\bibinfo {title} {Models for choice-reaction time},\ }\href {https://doi.org/10.1007/BF02289729} {\bibfield  {journal} {\bibinfo  {journal} {Psychometrika}\ }\textbf {\bibinfo {volume} {25}},\ \bibinfo {pages} {251} (\bibinfo {year} {1960})}\BibitemShut {NoStop}%
\bibitem [{\citenamefont {Zhang}\ and\ \citenamefont {Rowe}(2014)}]{Zhang:2014:SpeedAccuracy}%
  \BibitemOpen
  \bibfield  {author} {\bibinfo {author} {\bibfnamefont {J.}~\bibnamefont {Zhang}}\ and\ \bibinfo {author} {\bibfnamefont {J.~B.}\ \bibnamefont {Rowe}},\ }\bibfield  {title} {\bibinfo {title} {Dissociable mechanisms of speed-accuracy tradeoff during visual perceptual learning are revealed by a hierarchical drift-diffusion model},\ }\href {https://doi.org/10.3389/fnins.2014.00069} {\bibfield  {journal} {\bibinfo  {journal} {Front. Neurosci.}\ }\textbf {\bibinfo {volume} {8}} (\bibinfo {year} {2014})}\BibitemShut {NoStop}%
\bibitem [{\citenamefont {Erker}\ \emph {et~al.}(2017)\citenamefont {Erker}, \citenamefont {Mitchison}, \citenamefont {Silva}, \citenamefont {Woods}, \citenamefont {Brunner},\ and\ \citenamefont {Huber}}]{Erker:2017:QClockTUR}%
  \BibitemOpen
  \bibfield  {author} {\bibinfo {author} {\bibfnamefont {P.}~\bibnamefont {Erker}}, \bibinfo {author} {\bibfnamefont {M.~T.}\ \bibnamefont {Mitchison}}, \bibinfo {author} {\bibfnamefont {R.}~\bibnamefont {Silva}}, \bibinfo {author} {\bibfnamefont {M.~P.}\ \bibnamefont {Woods}}, \bibinfo {author} {\bibfnamefont {N.}~\bibnamefont {Brunner}},\ and\ \bibinfo {author} {\bibfnamefont {M.}~\bibnamefont {Huber}},\ }\bibfield  {title} {\bibinfo {title} {Autonomous quantum clocks: Does thermodynamics limit our ability to measure time?},\ }\href {https://doi.org/10.1103/PhysRevX.7.031022} {\bibfield  {journal} {\bibinfo  {journal} {Phys. Rev. X}\ }\textbf {\bibinfo {volume} {7}},\ \bibinfo {pages} {031022} (\bibinfo {year} {2017})}\BibitemShut {NoStop}%
\bibitem [{\citenamefont {Brandner}\ \emph {et~al.}(2018)\citenamefont {Brandner}, \citenamefont {Hanazato},\ and\ \citenamefont {Saito}}]{Brandner:2018:Transport}%
  \BibitemOpen
  \bibfield  {author} {\bibinfo {author} {\bibfnamefont {K.}~\bibnamefont {Brandner}}, \bibinfo {author} {\bibfnamefont {T.}~\bibnamefont {Hanazato}},\ and\ \bibinfo {author} {\bibfnamefont {K.}~\bibnamefont {Saito}},\ }\bibfield  {title} {\bibinfo {title} {Thermodynamic bounds on precision in ballistic multiterminal transport},\ }\href {https://doi.org/10.1103/PhysRevLett.120.090601} {\bibfield  {journal} {\bibinfo  {journal} {Phys. Rev. Lett.}\ }\textbf {\bibinfo {volume} {120}},\ \bibinfo {pages} {090601} (\bibinfo {year} {2018})}\BibitemShut {NoStop}%
\bibitem [{\citenamefont {Carollo}\ \emph {et~al.}(2019)\citenamefont {Carollo}, \citenamefont {Jack},\ and\ \citenamefont {Garrahan}}]{Carollo:2019:QuantumLDP}%
  \BibitemOpen
  \bibfield  {author} {\bibinfo {author} {\bibfnamefont {F.}~\bibnamefont {Carollo}}, \bibinfo {author} {\bibfnamefont {R.~L.}\ \bibnamefont {Jack}},\ and\ \bibinfo {author} {\bibfnamefont {J.~P.}\ \bibnamefont {Garrahan}},\ }\bibfield  {title} {\bibinfo {title} {Unraveling the large deviation statistics of {Markovian} open quantum systems},\ }\href {https://doi.org/10.1103/PhysRevLett.122.130605} {\bibfield  {journal} {\bibinfo  {journal} {Phys. Rev. Lett.}\ }\textbf {\bibinfo {volume} {122}},\ \bibinfo {pages} {130605} (\bibinfo {year} {2019})}\BibitemShut {NoStop}%
\bibitem [{\citenamefont {Liu}\ and\ \citenamefont {Segal}(2019)}]{Liu:2019:QTUR}%
  \BibitemOpen
  \bibfield  {author} {\bibinfo {author} {\bibfnamefont {J.}~\bibnamefont {Liu}}\ and\ \bibinfo {author} {\bibfnamefont {D.}~\bibnamefont {Segal}},\ }\bibfield  {title} {\bibinfo {title} {Thermodynamic uncertainty relation in quantum thermoelectric junctions},\ }\href {https://doi.org/10.1103/PhysRevE.99.062141} {\bibfield  {journal} {\bibinfo  {journal} {Phys. Rev. E}\ }\textbf {\bibinfo {volume} {99}},\ \bibinfo {pages} {062141} (\bibinfo {year} {2019})}\BibitemShut {NoStop}%
\bibitem [{\citenamefont {Guarnieri}\ \emph {et~al.}(2019)\citenamefont {Guarnieri}, \citenamefont {Landi}, \citenamefont {Clark},\ and\ \citenamefont {Goold}}]{Guarnieri:2019:QTURPRR}%
  \BibitemOpen
  \bibfield  {author} {\bibinfo {author} {\bibfnamefont {G.}~\bibnamefont {Guarnieri}}, \bibinfo {author} {\bibfnamefont {G.~T.}\ \bibnamefont {Landi}}, \bibinfo {author} {\bibfnamefont {S.~R.}\ \bibnamefont {Clark}},\ and\ \bibinfo {author} {\bibfnamefont {J.}~\bibnamefont {Goold}},\ }\bibfield  {title} {\bibinfo {title} {Thermodynamics of precision in quantum nonequilibrium steady states},\ }\href {https://doi.org/10.1103/PhysRevResearch.1.033021} {\bibfield  {journal} {\bibinfo  {journal} {Phys. Rev. Research}\ }\textbf {\bibinfo {volume} {1}},\ \bibinfo {pages} {033021} (\bibinfo {year} {2019})}\BibitemShut {NoStop}%
\bibitem [{\citenamefont {Saryal}\ \emph {et~al.}(2019)\citenamefont {Saryal}, \citenamefont {Friedman}, \citenamefont {Segal},\ and\ \citenamefont {Agarwalla}}]{Saryal:2019:TUR}%
  \BibitemOpen
  \bibfield  {author} {\bibinfo {author} {\bibfnamefont {S.}~\bibnamefont {Saryal}}, \bibinfo {author} {\bibfnamefont {H.~M.}\ \bibnamefont {Friedman}}, \bibinfo {author} {\bibfnamefont {D.}~\bibnamefont {Segal}},\ and\ \bibinfo {author} {\bibfnamefont {B.~K.}\ \bibnamefont {Agarwalla}},\ }\bibfield  {title} {\bibinfo {title} {Thermodynamic uncertainty relation in thermal transport},\ }\href {https://link.aps.org/doi/10.1103/PhysRevE.100.042101} {\bibfield  {journal} {\bibinfo  {journal} {Phys. Rev. E}\ }\textbf {\bibinfo {volume} {100}},\ \bibinfo {pages} {042101} (\bibinfo {year} {2019})}\BibitemShut {NoStop}%
\bibitem [{\citenamefont {Hasegawa}(2020)}]{Hasegawa:2020:QTURPRL}%
  \BibitemOpen
  \bibfield  {author} {\bibinfo {author} {\bibfnamefont {Y.}~\bibnamefont {Hasegawa}},\ }\bibfield  {title} {\bibinfo {title} {Quantum thermodynamic uncertainty relation for continuous measurement},\ }\href {https://doi.org/10.1103/PhysRevLett.125.050601} {\bibfield  {journal} {\bibinfo  {journal} {Phys. Rev. Lett.}\ }\textbf {\bibinfo {volume} {125}},\ \bibinfo {pages} {050601} (\bibinfo {year} {2020})}\BibitemShut {NoStop}%
\bibitem [{\citenamefont {Hasegawa}(2021)}]{Hasegawa:2020:TUROQS}%
  \BibitemOpen
  \bibfield  {author} {\bibinfo {author} {\bibfnamefont {Y.}~\bibnamefont {Hasegawa}},\ }\bibfield  {title} {\bibinfo {title} {Thermodynamic uncertainty relation for general open quantum systems},\ }\href {https://doi.org/10.1103/PhysRevLett.126.010602} {\bibfield  {journal} {\bibinfo  {journal} {Phys. Rev. Lett.}\ }\textbf {\bibinfo {volume} {126}},\ \bibinfo {pages} {010602} (\bibinfo {year} {2021})}\BibitemShut {NoStop}%
\bibitem [{\citenamefont {Kalaee}\ \emph {et~al.}(2021)\citenamefont {Kalaee}, \citenamefont {Wacker},\ and\ \citenamefont {Potts}}]{Kalaee:2021:QTURPRE}%
  \BibitemOpen
  \bibfield  {author} {\bibinfo {author} {\bibfnamefont {A.~A.~S.}\ \bibnamefont {Kalaee}}, \bibinfo {author} {\bibfnamefont {A.}~\bibnamefont {Wacker}},\ and\ \bibinfo {author} {\bibfnamefont {P.~P.}\ \bibnamefont {Potts}},\ }\bibfield  {title} {\bibinfo {title} {Violating the thermodynamic uncertainty relation in the three-level maser},\ }\href {https://doi.org/10.1103/PhysRevE.104.L012103} {\bibfield  {journal} {\bibinfo  {journal} {Phys. Rev. E}\ }\textbf {\bibinfo {volume} {104}},\ \bibinfo {pages} {L012103} (\bibinfo {year} {2021})}\BibitemShut {NoStop}%
\bibitem [{\citenamefont {Monnai}(2022)}]{Monnai:2022:QTUR}%
  \BibitemOpen
  \bibfield  {author} {\bibinfo {author} {\bibfnamefont {T.}~\bibnamefont {Monnai}},\ }\bibfield  {title} {\bibinfo {title} {Thermodynamic uncertainty relation for quantum work distribution: Exact case study for a perturbed oscillator},\ }\href {https://doi.org/10.1103/PhysRevE.105.034115} {\bibfield  {journal} {\bibinfo  {journal} {Phys. Rev. E}\ }\textbf {\bibinfo {volume} {105}},\ \bibinfo {pages} {034115} (\bibinfo {year} {2022})}\BibitemShut {NoStop}%
\bibitem [{\citenamefont {Hasegawa}(2023)}]{Hasegawa:2023:BulkBoundaryBoundNC}%
  \BibitemOpen
  \bibfield  {author} {\bibinfo {author} {\bibfnamefont {Y.}~\bibnamefont {Hasegawa}},\ }\bibfield  {title} {\bibinfo {title} {Unifying speed limit, thermodynamic uncertainty relation and {Heisenberg} principle via bulk-boundary correspondence},\ }\href {https://doi.org/10.1038/s41467-023-38074-8} {\bibfield  {journal} {\bibinfo  {journal} {Nat. Commun.}\ }\textbf {\bibinfo {volume} {14}},\ \bibinfo {pages} {2828} (\bibinfo {year} {2023})}\BibitemShut {NoStop}%
\bibitem [{\citenamefont {Nishiyama}\ and\ \citenamefont {Hasegawa}(2024)}]{Nishiyama:2024:OpenQuantumRURJPA}%
  \BibitemOpen
  \bibfield  {author} {\bibinfo {author} {\bibfnamefont {T.}~\bibnamefont {Nishiyama}}\ and\ \bibinfo {author} {\bibfnamefont {Y.}~\bibnamefont {Hasegawa}},\ }\bibfield  {title} {\bibinfo {title} {Tradeoff relations in open quantum dynamics via {Robertson}, {Maccone}-{Pati}, and {Robertson}-{Schr{\"o}dinger} uncertainty relations},\ }\href {https://doi.org/10.1088/1751-8121/ad79cd} {\bibfield  {journal} {\bibinfo  {journal} {J. Phys. A: Math. Theor.}\ }\textbf {\bibinfo {volume} {57}},\ \bibinfo {pages} {415301} (\bibinfo {year} {2024})}\BibitemShut {NoStop}%
\bibitem [{\citenamefont {Hasegawa}\ and\ \citenamefont {Nishiyama}(2024)}]{Hasegawa:2024:ConcentrationIneqPRL}%
  \BibitemOpen
  \bibfield  {author} {\bibinfo {author} {\bibfnamefont {Y.}~\bibnamefont {Hasegawa}}\ and\ \bibinfo {author} {\bibfnamefont {T.}~\bibnamefont {Nishiyama}},\ }\bibfield  {title} {\bibinfo {title} {Thermodynamic concentration inequalities and trade-off relations},\ }\href {https://doi.org/10.1103/PhysRevLett.133.247101} {\bibfield  {journal} {\bibinfo  {journal} {Phys. Rev. Lett.}\ }\textbf {\bibinfo {volume} {133}},\ \bibinfo {pages} {247101} (\bibinfo {year} {2024})}\BibitemShut {NoStop}%
\bibitem [{\citenamefont {Prech}\ \emph {et~al.}(2025)\citenamefont {Prech}, \citenamefont {Potts},\ and\ \citenamefont {Landi}}]{Prech:2025:CoherenceQTUR}%
  \BibitemOpen
  \bibfield  {author} {\bibinfo {author} {\bibfnamefont {K.}~\bibnamefont {Prech}}, \bibinfo {author} {\bibfnamefont {P.~P.}\ \bibnamefont {Potts}},\ and\ \bibinfo {author} {\bibfnamefont {G.~T.}\ \bibnamefont {Landi}},\ }\bibfield  {title} {\bibinfo {title} {Role of quantum coherence in kinetic uncertainty relations},\ }\href {https://doi.org/10.1103/PhysRevLett.134.020401} {\bibfield  {journal} {\bibinfo  {journal} {Phys. Rev. Lett.}\ }\textbf {\bibinfo {volume} {134}},\ \bibinfo {pages} {020401} (\bibinfo {year} {2025})}\BibitemShut {NoStop}%
\bibitem [{\citenamefont {Risken}(1989)}]{Risken:1989:FPEBook}%
  \BibitemOpen
  \bibfield  {author} {\bibinfo {author} {\bibfnamefont {H.}~\bibnamefont {Risken}},\ }\href@noop {} {\emph {\bibinfo {title} {The {Fokker}--{Planck} Equation: Methods of Solution and Applications}}},\ \bibinfo {edition} {2nd}\ ed.\ (\bibinfo  {publisher} {Springer},\ \bibinfo {year} {1989})\BibitemShut {NoStop}%
\bibitem [{\citenamefont {Hasegawa}(2025)}]{Data:2025:TEUR}%
  \BibitemOpen
  \bibfield  {author} {\bibinfo {author} {\bibfnamefont {Y.}~\bibnamefont {Hasegawa}},\ }\href@noop {} {\bibinfo {title} {Thermodynamic entropic uncertainty relation}},\ \bibinfo {howpublished} {\url{https://github.com/yoshihiko-hasegawa/ThermodynamicEntropicUncertaintyRelation}} (\bibinfo {year} {2025}),\ \bibinfo {note} {{GitHub} repository}\BibitemShut {NoStop}%
\end{thebibliography}
\end{document}


\title{Supplementary Material for\\ ``Thermodynamic Entropic Uncertainty Relation''}

\author{Yoshihiko Hasegawa}
\email{hasegawa@biom.t.u-tokyo.ac.jp}
\affiliation{Department of Information and Communication Engineering, Graduate
School of Information Science and Technology, The University of Tokyo,
Tokyo 113-8656, Japan}

\author{Tomohiro Nishiyama}
\email{htam0ybboh@gmail.com}
\affiliation{Independent Researcher, Tokyo 206-0003, Japan}

\maketitle

\section{General observables}

In the main text, we have focused on observables that satisfy the time-reversal antisymmetry condition given by Eq.~\timeUreversal{}. 
In this section, we relax this condition and introduce another optimization framework. 
Here, in order to provide a clearer perspective, we first embed the classical system into a quantum system.
We define a quantum state $\ket{\Gamma}$ corresponding to the trajectory $\Gamma$. 
As in usual quantum mechanics,
the bra vector is defined as
$\bra{\Gamma}=\ket{\Gamma}^\dagger$. 
It satisfies orthogonality
\begin{align}
    \braket{\Gamma^\prime|\Gamma} = \delta(\Gamma = \Gamma^\prime),
    \label{eq:ortho_condition}
\end{align}
where $\delta(\bullet)$ is $1$ if the argument is true and $0$ otherwise. 
Furthermore, it is possible to define a density operator related to the probability of the trajectory $\mathcal{P}(\Gamma)$. This density operator is expressed as
\begin{align}
    \rho=\sum_{\Gamma}\mathcal{P}(\Gamma)\ket{\Gamma}\bra{\Gamma}.
    \label{eq:rho_def}
\end{align}
Since $\rho$ defined in Eq.~\eqref{eq:rho_def} is a standard density operator, it satisfies $\mathrm{Tr}[\rho] = 1$. 
We introduce the quantum channel $\mathcal{R}(\rho)$ corresponding to the time-reversal operation
\begin{align}
    \mathcal{R}(\rho)\equiv R \rho R = R \rho R^\dagger,
    \label{eq:channel_R}
\end{align}
where the operator $R$ is defined by
\begin{align}
    R\equiv\sum_{\Gamma}\ket{\Gamma^{\dagger}}\bra{\Gamma}.
    \label{eq:time_rev_R_def}
\end{align}
Here, $\Gamma^\dagger$ is the time-reversal of $\Gamma$,
as defined in the main text. 
Projection-valued measure corresponding to the observable $\Phi(\Gamma)$ is given by
\begin{align}
    M_{\phi}\equiv\sum_{\Gamma}\delta(\Phi(\Gamma)=\phi)\ket{\Gamma}\bra{\Gamma}.
    \label{eq:Mx_POVM_def}
\end{align}
Using $M_\phi$, the probability of measuring $\Phi(\Gamma)=\phi$ can be calculated by
\begin{align}
    P(\Phi=\phi)=\mathrm{Tr}[M_{\phi}\rho].
    \label{eq:PPhi_by_PVM}
\end{align}
With Eqs.~\eqref{eq:rho_def} and \eqref{eq:Mx_POVM_def}, the classical stochastic thermodynamic system can be embedded into the corresponding quantum dynamics. 
We can calculate the probability for the time-reversed state as follows:
\begin{align}
    P(\Phi(\Gamma^{\dagger})=\phi)=P(\Phi(\Gamma)=-\phi)=\mathrm{Tr}[M_{\phi}\mathcal{R}(\rho)]=\mathrm{Tr}[RM_{\phi}R\rho]=\mathrm{Tr}[M_{\phi}^{\prime}\rho],
    \label{eq:Mx_RrhoR}
\end{align}
where $M_{\phi}^{\prime}\equiv RM_{\phi}R$. 
Since $\sum_\phi M_\phi = \mathbb{I}$, it is easy to show that $\sum_{\phi}M_{\phi}^{\prime}=\mathbb{I}$, indicating that $M^\prime_\phi$ also constitutes a proper projection-valued measure. 
Let us investigate the relationship between $M_\phi$ and $M_\phi^\prime$. 
Their product is given by
\begin{align}
    M_{\phi}M_{\phi}^{\prime}=\sum_{\Gamma}\delta(\Phi(\Gamma)=\phi)\delta(\Phi(\Gamma^{\dagger})=\phi)\ket{\Gamma}\bra{\Gamma}.
    \label{eq:Mphi_Mphidash_prod}
\end{align}
Suppose that $\Phi$ does not include $\Phi=0$. 
If the observable $\Phi(\Gamma)$ satisfies the time-reversal antisymmetry given in Eq.~\timeUreversal{}, $M_\phi$ and $M_\phi^\prime$ are orthogonal:
\begin{align}
    M_{\phi}M_{\phi}^{\prime}=0.
    \label{eq:Mx_orthogonal}
\end{align}
When $M_\phi$ and $M_\phi^\prime$ are orthogonal, $M_\phi + M^\prime_\phi$ is a projector. By using the H\"older inequality, the following relation holds:
\begin{align}
    \mathrm{Tr}[(M_{\phi}+M_{\phi}^{\prime})\rho]\le\left\Vert M_{\phi}+M_{\phi}^{\prime}\right\Vert _{\mathrm{op}}\left\Vert \rho\right\Vert _{\mathrm{tr}}=1.
    \label{eq:MMprime_upperbound}
\end{align}
where $\left\Vert \bullet\right\Vert _{\mathrm{op}}$ and $\left\Vert \bullet\right\Vert _{\mathrm{tr}}$ are operator and trace norms, respectively. 
In a general situation, the observable may not satisfy the time-reversal antisymmetry condition. In such a case, the following relation holds:
\begin{align}
    \mathrm{Tr}[(M_{\phi}+M_{\phi}^{\prime})\rho]&=\mathrm{Tr}[(M_{\phi}+M_{\phi}^{\prime}-M_{\phi}M_{\phi}^{\prime}+M_{\phi}M_{\phi}^{\prime})\rho]\nonumber\\
    &\le1+\mathrm{Tr}[M_{\phi}M_{\phi}^{\prime}\rho],
    \label{eq:Mphi_Mphiprime_sum}
\end{align}
where we used the fact that $M_{\phi}+M_{\phi}^{\prime}-M_{\phi}M_{\phi}^{\prime}$ is a projector. 
From Eq.~\eqref{eq:Mphi_Mphiprime_sum}, the following relation holds:
\begin{align}
    P(\phi)+P(-\phi)\le1+\alpha_{\phi},
    \label{eq:Phi_Phineg_sum}
\end{align}
where $\alpha_{\phi}\equiv\mathrm{Tr}[M_{\phi}M_{\phi}^{\prime}\rho]$. 
$\alpha_\phi$ quantifies the overlap between $M_\phi$ and $M_\phi^\prime$:
\begin{align}
    \alpha_{\phi}&=\sum_{\Gamma}\delta(\Phi(\Gamma)=\phi)\delta(\Phi(\Gamma^{\dagger})=\phi)\mathcal{P}(\Gamma),
    \label{eq:alpha_phi}
\end{align}
which denotes the sum of the probabilities of the set of trajectories corresponding to the value $\phi$ for which the time-reversed trajectories also take the same value $\phi$.
If the observable $\Phi$ satisfies time-reversal antisymmetry (i.e., $\Phi(\Gamma^\dagger) = -\Phi(\Gamma)$), then trajectories with the same value $\phi$ and their time-reversed counterparts cannot coexist, $\alpha_\phi = 0$. 
On the other hand, if this symmetry is broken (meaning that $\Phi(\Gamma^\dagger)$ and $\Phi(\Gamma)$ can take the same value), then $\alpha_\phi > 0$.
As shown in the main text [Eq.~\SigmaUmonotoniciy{}], we have shown
\begin{align}
    \Sigma+H[P(\Phi)]\ge-\sum_{\phi}P(\phi)\ln P(-\phi).
    \label{eq:Sigma_H_sum}
\end{align}
Therefore, the bound for $\Sigma + H[\Phi(\phi)]$ can be formulated as a minimization of $-\sum_\phi P(\phi) \ln P(-\phi)$ under the constraint given by Eq.~\eqref{eq:Phi_Phineg_sum}.
Then, redefining the variables $q_\phi \equiv P(\phi)$ and $q_\phi^\prime \equiv P(-\phi)$, the minimization problem is 
\begin{align}
    \underset{q_{\phi},q_{\phi}^{\prime}}{\mathrm{minimize}}-\sum_{\phi}q_{\phi}\ln q_{\phi}^{\prime},
    \label{eq:objective_function}
\end{align}
such that
\begin{align}
    0&\le q_{\phi}+q_{\phi}^{\prime}\le1+\alpha_{\phi},\label{eq:constraint1}\\
    0&\le q_{\phi}\le1,\label{eq:constraint2}\\
    0&\le q_{\phi}^{\prime}\le1,\label{eq:constraint3}\\
    1&=\sum_{\phi}q_{\phi},\label{eq:constraint4}\\
    1&=\sum_{\phi}q_{\phi}^{\prime}.\label{eq:constraint5}
\end{align}
The conditions of Eqs.~\eqref{eq:constraint1}--\eqref{eq:constraint5} must be satisfied for all $\phi$. 
We aim to optimize Eq.~\eqref{eq:objective_function}, subject to the constraints specified in Eqs.~\eqref{eq:constraint1} through~\eqref{eq:constraint5}. This task presents a nonlinear constrained optimization problem. To find the minimum, the solution must satisfy the Karush-Kuhn-Tucker (KKT) conditions \cite{Boyd:2004:ConvexOptimBook}.
When $\alpha_\phi = 1$, the optimization becomes trivial because both $q_\phi$ and $q_\phi^\prime$ can assume any probability values, resulting in the minimum of $0$. However, if $\alpha_\phi < 1$, $q_\phi$ and $q_\phi^\prime$ cannot take arbitrary values, making the optimization problem more complex, making the minimum nontrivial.
As mentioned earlier, when the observable $\Phi(\Gamma)$ satisfies the time-reversal antisymmetry (and $P(\Phi=0)=0$), $\alpha_\phi = 0$.
Therefore, employing the values of $\alpha_\phi > 0$,
we can consider cases where the observable partially violates the time-reversal antisymmetry condition. 

To minimize Eq.~\eqref{eq:objective_function} under the constraints Eqs.~\eqref{eq:constraint1}--\eqref{eq:constraint5}, 
let $\mathcal{L}$ be
\begin{align}
    \mathcal{L}=-\sum_{\phi}q_{\phi}\ln q_{\phi}^{\prime}+\sum_{\phi}\mu_{\phi}\left(q_{\phi}+q_{\phi}^{\prime}-1-\alpha_{\phi}\right)+\lambda_{1}\left(1-\sum_{\phi}q_{\phi}\right)+\lambda_{2}\left(1-\sum_{\phi}q_{\phi}^{\prime}\right),
    \label{eq:ineq_eq_Lagrangian_def}
\end{align}
where $\lambda_1$, $\lambda_2$, and $\mu_\phi$ denote the Lagrange multipliers. 
Based on the KKT conditions, the optimal solution must meet these specified criteria:
\begin{align}
    \frac{\partial\mathcal{L}}{\partial q_{\phi}}&=-\ln q_{\phi}^{\prime}+\mu_{\phi}-\lambda_{1}=0,\label{eq:KKT1}\\\frac{\partial\mathcal{L}}{\partial q_{\phi}^{\prime}}&=-\frac{q_{\phi}}{q_{\phi}^{\prime}}+\mu_{\phi}-\lambda_{2}=0,\label{eq:KKT2}\\\mu_{\phi}\left(q_{\phi}+q_{\phi}^{\prime}-1-\alpha_\phi\right)&=0,\label{eq:KKT3}\\\mu_{\phi}&\ge0.\label{eq:KKT4}
\end{align}
For a general $\alpha_\phi$, the bound becomes
\begin{align}
\Sigma + H[P(\Phi)] \ge \mathcal{M},
\label{eq:TEUR_general_case}
\end{align}
where $\mathcal{M}$ is the minimum value of Eq.~\eqref{eq:objective_function}.
It is not feasible to analytically find the minimum solution in this scenario. Therefore, we must use numerical methods to determine $\mathcal{M}$.

When $\Phi(\Gamma)$ is a binary function and satisfies the antisymmetry condition under the time-reversal.
In this case, $\alpha_\phi = 0$ and we can solve the optimization problem analytically to obtain
\begin{align}
    \Sigma + H[P(\Phi)] \ge \ln 2,
    \label{eq:GTEUR_ln2}
\end{align}
which recovers the bound derived in Eq.~\mainUresultUbinary{} in the main text.